\documentclass[aps,prb,onecolumn,superscriptaddress,amsmath,amssymb,floatfix]{revtex4-2}
\usepackage{graphicx}
\usepackage{bm}
\usepackage{booktabs}
\usepackage{enumitem}
\usepackage{xcolor}
\usepackage[colorlinks,linkcolor=blue,citecolor=red,urlcolor=blue]{hyperref}
\usepackage[caption=false]{subfig}
 
\setlist{itemsep=2pt,topsep=4pt}

\newcommand{\bR}{\mathbf R}
\newcommand{\bk}{\mathbf k}
\newcommand{\dk}{\delta\mathbf k}
\newcommand{\Rh}{\hat{\mathbf R}}
\newcommand{\up}{\uparrow}
\newcommand{\dn}{\downarrow}
\newcommand{\Jc}{\mathcal J}
\newcommand{\eF}{\varepsilon_F}
\newcommand{\Jp}{J_\perp}
\newcommand{\bS}{\mathbf S}

\newcommand{\mref}[1]{\eqref{#1}}     
\newcommand{\sref}[1]{\eqref{#1}}     
 
\newsavebox{\figtwobox}
\newsavebox{\figthreebox}

\begin{document}

\title{Current-driven chiral RKKY interaction in altermagnets}
\author{A.~Kundu}
\affiliation{Asia Pacific Center for Theoretical Physics, POSTECH, Pohang 37673, Korea.}
\begin{abstract}
The Dzyaloshinskii--Moriya (DM) interaction between magnetic moments usually requires spin--orbit
coupling and broken inversion symmetry. We show that in a spin-split metal without spin--orbit
coupling a charge current generates a DM interaction between localized spins. Using Keldysh Green's
functions (GFs) with a drifting Fermi sea, we compute the Ruderman--Kittel--Kasuya--Yosida (RKKY)
interaction between two magnetic impurities in a two-dimensional $d$-wave altermagnet with an
additional exchange splitting. 
To linear order in the current, the correction is a Fermi-surface term,
and all its components are obtained in closed form in terms of Bessel
functions, valid for arbitrary values of the two independent spin
splittings, $\beta$ and $J_\perp$.
The induced DM vector points along the spin-quantization axis, which in
the present model is the out-of-plane direction $\hat{\mathbf{z}}$. At
long distance it decays as $1/R$, one power more slowly than the
equilibrium coupling.
In the altermagnet it vanishes for bonds along the nodal directions, where the two spin bands are degenerate, and,
for a diagonal current, has opposite signs on bonds along the
two crystal axes. 
This $d$-wave signature reverses with the N\'eel vector and with the current, and
distinguishes the altermagnet from a ferromagnet. 
The current also generates symmetric couplings
that are odd in the impurity separation. In
low-density two-dimensional conductors, the DM coupling can reach a few percent of the
equilibrium exchange.
\end{abstract}

\maketitle

\section{Introduction}
\label{sec:intro}
 
The Ruderman--Kittel--Kasuya--Yosida (RKKY)
interaction~\cite{Ruderman1954,Kasuya1956,Yosida1957}, the indirect exchange between localized
moments mediated by conduction electrons, governs the magnetism of dilute magnetic alloys and the
coupling of magnetic layers and adatoms, and it can now be mapped atom by atom on metal
surfaces~\cite{Zhou2010,Khajetoorians2016}. In a two-dimensional electron gas it oscillates with
twice the Fermi wave number and decays as $1/R^2$ with the separation $R$~\cite{PhysRevB.58.3584}.
Because its form reflects the electronic structure of the host, the RKKY interaction has been
studied in a wide range of materials, including graphene~\cite{PhysRevB.84.115119, Sherafati2011},
topological insulators~\cite{Liu2009,PhysRevLett.106.097201}, and Dirac and Weyl
semimetals~\cite{Chang2015,Hosseini2015}.
 
Besides symmetric exchange, the coupling between two spins can contain an antisymmetric
Dzyaloshinskii--Moriya (DM) term $\mathbf D\cdot(\bS_1\times\bS_2)$~\cite{Dzyaloshinsky1958,Moriya1960},
which favors noncollinear, chiral arrangements and stabilizes chiral domain walls and magnetic
skyrmions~\cite{Fert2017}. In equilibrium, a DM interaction requires broken inversion symmetry and
spin--orbit coupling. For the RKKY coupling it originates from spin--orbit
scattering~\cite{PhysRevLett.44.1538} or from Rashba coupling of the conduction
electrons~\cite{PhysRevB.69.121303,Bouaziz_2017}, and it has been tailored for individual pairs of
adatoms~\cite{Khajetoorians2016}. Electrical control of the DM interaction is therefore usually tied
to the control of spin--orbit coupling, for example by a gate voltage
that changes the interfacial DM interaction in heavy-metal/ferromagnet
films~\cite{Srivastava2018,Koyama2018}, or to
special band structures such as those of magnetic Weyl semimetals~\cite{PhysRevB.108.134428}. In
spin--orbit-coupled magnets the DM interaction is closely linked to spin
currents~\cite{Kikuchi2016}, and electric currents have been predicted and reported to modify it in
magnetic multilayers~\cite{Freimuth2020,Karnad2018}.
 
Altermagnets~\cite{smejkal2022beyond,smejkal2022emerging} offer a different route to spin-split
electronic bands. In these collinear magnets the opposite-spin sublattices are related by a
rotation rather than by a translation or inversion, which produces a momentum-dependent spin
splitting of nonrelativistic origin while the net magnetization vanishes. Altermagnetic spin
splitting has been observed in MnTe~\cite{Krempasky2024_Nature,Amin2024}, in
CrSb~\cite{Reimers2024}, and in the layered metals KV$_2$Se$_2$O~\cite{Jiang2025} and
Rb$_{1-\delta}$V$_2$Te$_2$O~\cite{Zhang2025}, which show $d$-wave splitting above room temperature.
Because the splitting originates from exchange, spin remains a good quantum number, and a charge
current carries a spin current whose polarization depends on the direction of flow; this is the
spin-splitter effect~\cite{GonzalezHernandez2021,bose2022tilted}. The RKKY interaction in
altermagnets has so far been studied in
equilibrium~\cite{PhysRevB.110.054427,Lee2025,Yarmohammadi_2025,Kundu2025}. It acquires a
characteristic anisotropy, but a DM term appears only when Rashba spin--orbit coupling is
added~\cite{PhysRevB.110.054427,Yarmohammadi_2025,Kundu2025}.
 
Out of equilibrium, the indirect exchange is no longer fixed by the equilibrium spin
susceptibility. Theoretical study of Nonequilibrium RKKY couplings have been computed across biased tunnel
junctions~\cite{Schwabe1996} and for spins coupled to spin-polarized substrates and
leads~\cite{Fransson2010,Fransson2014}, charge and spin currents have been shown to change the
dynamics of two localized spins qualitatively~\cite{Onoda2006}, and a bias voltage or light
irradiation can make the RKKY interaction
nonreciprocal~\cite{Cherukupalli2026,Hanai2025}.
Whether a charge current can generate a chiral exchange interaction in a metal whose bands are spin split but which has no spin--orbit coupling, and what spatial structure such an interaction acquires in an altermagnet, has so far not been addressed.
 
In this paper we answer this question for a minimal model, a two-dimensional $d$-wave altermagnet
with an additional uniform exchange splitting $\Jp$, which interpolates between the pure
altermagnet and a ferromagnet. We combine the Keldysh expression for the static exchange coupling
with a Boltzmann drift of the Fermi sea. 
To linear order in the current, the current-induced correction is a
pure Fermi-surface term, and all its components follow in closed form
in terms of Bessel functions, exactly in both spin splittings.
We find that the current generates a DM
interaction $\delta D_z(\bS_1\times\bS_2)_z$ without spin--orbit coupling. It decays as $1/R$,
one power more slowly than the equilibrium coupling, vanishes in a paramagnet, and in the
altermagnet it has a $d$-wave angular dependence that distinguishes it from the ferromagnetic case.
The current also generates symmetric couplings that are odd in $\bR$ and cannot be derived from a
spin Hamiltonian. We estimate the size of these effects and identify candidate materials.
 
The paper is organized as follows. Section~\ref{sec:model} introduces the model and the Keldysh
expression for the exchange coupling. Section~\ref{sec-equill-rkky} derives the equilibrium couplings,
and Sec.~\ref{sec: non-equillibrium-RKKY} the current-induced couplings, their symmetry, their long-distance
forms, and their magnitudes. Section~\ref{sec:figures} illustrates the results numerically.
Section~\ref{sec:discussion} discusses the assumptions, the relation to previous work, and the
experimental prospects, and Sec.~\ref{sec:summary} summarizes our results.

\section{Model, conventions and assumptions}
\label{sec:model}
We consider a minimal effective-mass model of a two-dimensional
$d$-wave altermagnet with an additional exchange (Zeeman-like)
splitting,
\begin{equation}
H_{\bk}=\frac{\hbar^2k^2}{2m}\,\sigma_0+\big[\Jp+\beta\,(k_x^2-k_y^2)\big]\,\sigma_z,
\label{eq:H}
\end{equation}
with $\sigma=+1\,(\up)$, $-1\,(\dn)$. 
The band energy is $\varepsilon_\sigma(\bk)=\hbar^2k_x^2/2m_{x\sigma}+\hbar^2k_y^2/2m_{y\sigma}+\sigma\Jp$, with spin-dependent masses
\begin{equation}
\frac{1}{m_{x\sigma}}=\frac{1}{m}(1+\sigma b),\qquad \frac{1}{m_{y\sigma}}=\frac{1}{m}(1-\sigma b),\qquad
b\equiv\frac{2m\beta}{\hbar^2}=\frac{\beta k_F^2}{\eF},\quad |b|<1 .
\label{eq:masses}
\end{equation}
 The Fermi surface (FS) of each spin band is an ellipse, and a $90^\circ$ rotation interchanges the two ($\up\leftrightarrow\dn$).
The $s$--$d$ coupling between the conduction electrons and the spins $\bS_i$ of two magnetic impurities separated by $\bR=R(\cos\phi,\sin\phi)$ is,
\begin{equation}
H_{\rm int}=\Jc\sum_{i=1,2}\bm\sigma\cdot\bS_i\,\delta(\mathbf r-\bR_i),
\label{eq:Hsd}
\end{equation}
where $\Jc$ is the local exchange coupling, and the electron spin density is $s^a(\mathbf r)=\psi^\dagger(\mathbf r)\sigma^a\psi(\mathbf r)$.
Impurity~2 couples to this density through
Eq.~\eqref{eq:Hsd}, which gives the indirect (RKKY) exchange $H_{\rm RKKY}=\sum_{ab}S_2^a\,J_{ab}(\bR)\,S_1^b$ with $
J_{ab}(\bR)=\Jc^2\,\chi^R_{ab}(\bR).$
Out of equilibrium, the exchange coupling is obtained from the Keldysh GFs~\cite{Keldysh1965,Rammer2007,HaugJauho2008,Kamenev2011,Stefanucci2013}.
\begin{equation}
J_{ab}(\bR)=-\frac{i\Jc^2}{2\pi}\int d\varepsilon\;{\rm Tr}\Big[\sigma_aG^R(\bR,\varepsilon)\,\sigma_bG^<(-\bR,\varepsilon)+\sigma_aG^<(\bR,\varepsilon)\,\sigma_bG^A(-\bR,\varepsilon)\Big].
\label{eq:Jneq}
\end{equation}
Since the current leaves $G^{R,A}(\bR,\varepsilon)$ unchanged and enters only through $G^<_\sigma(\bR,\varepsilon)=G^<_{\sigma,{\rm eq}}(\bR,\varepsilon)+\delta G^<_\sigma(\bR,\varepsilon)$, and Eq.~\eqref{eq:Jneq} is linear in $G^<$, the coupling splits exactly into the equilibrium RKKY tensor $J^{(0)}_{ab}$, built from $G^<_{\rm eq}$, and a current-induced term $\delta J_{ab}$, built from $\delta G^<$ [Eq.~\eqref{eq:Jsplit}]. As a consequence, the total RKKY interaction strength is,  $J_{ab}(\bR)=J^{(0)}_{ab}(\bR)+\delta J_{ab}(\bR)$ with 
\begin{equation}
\begin{array}{rcl}
J^{(0)}_{ab}(\bR)&=&-\dfrac{i\Jc^2}{2\pi}\displaystyle\int d\varepsilon\;{\rm Tr}\Big[\sigma_aG^R(\bR,\varepsilon)\,\sigma_bG^<_{\rm eq}(-\bR,\varepsilon)+\sigma_aG^<_{\rm eq}(\bR,\varepsilon)\,\sigma_bG^A(-\bR,\varepsilon)\Big],\\
\delta J_{ab}(\bR)&=&-\dfrac{i\Jc^2}{2\pi}\displaystyle\int d\varepsilon\;{\rm Tr}\Big[\sigma_aG^R(\bR,\varepsilon)\,\sigma_b\,\delta G^<(-\bR,\varepsilon)+\sigma_a\,\delta G^<(\bR,\varepsilon)\,\sigma_bG^A(-\bR,\varepsilon)\Big].
\end{array}
\label{eq:Jsplit}
\end{equation}
The real-space GFs are obtained by Fourier transforming the momentum-space GFs, which we carry out systematically below.

{\textit{GFs in k-space:}} 
The retarded GF corresponding to the Hamiltonian \eqref{eq:H} is,
\begin{equation}
G^{R}_\sigma(\bk,\varepsilon)=\dfrac{1}{\varepsilon-\varepsilon_\sigma(\bk)+i0}.
\end{equation}
The lesser GF can be written as, 
$G^<_\sigma(\bk,\varepsilon)=G^<_{\sigma,{\rm eq}}(\bk,\varepsilon)+\delta G^<_\sigma(\bk,\varepsilon)$.
The equilibrium part is
$G^<_{\sigma,{\rm eq}}(\bk,\varepsilon)=i f_0(\varepsilon)A_\sigma(\bk,\varepsilon)=-f_0(\varepsilon)\big[G^R_\sigma(\bk,\varepsilon)-G^A_\sigma(\bk,\varepsilon)\big]=2\pi i\,f_0(\varepsilon)\,\delta\big(\varepsilon-\varepsilon_\sigma(\bk)\big)$ 
where $A_\sigma(\bk,\varepsilon)$ is the spectral
function, $f_0$ is the Fermi function with chemical potential $\eF$.
The current-induced correction is,
\begin{equation}
\delta G^<_\sigma(\bk,\varepsilon)=i\big[-f_0'(\varepsilon)\big]\big[\dk\cdot\nabla_{\bk}\varepsilon_\sigma(\bk)\big]A_\sigma(\bk,\varepsilon),
\label{eq:gklesser}
\end{equation}
where $\dk=-e\tau\mathbf E/\hbar$ is the drift of the Fermi sea obtained from the Boltzmann transport equation (see SI).
 Since $-f_0'(\epsilon)$ is peaked at $\eF$,
$\delta G^<_\sigma(\bk,\varepsilon)\propto\dk$ is confined to the FS and describes its shift.
 In equilibrium
$\delta G^<_\sigma(\bk,\varepsilon)=0$, and Eq.~\eqref{eq:Jneq} reduces to the standard RKKY formula. 

\section{Zeroth order}
\label{sec-equill-rkky}
Inserting  $G^<_{\sigma,{\rm eq}}(\bR,\varepsilon)=i f_0(\varepsilon)A_\sigma(\bR,\varepsilon)=-2i\,f_0(\varepsilon)\,{\rm Im}\,G^R_\sigma(\bR,\varepsilon)$ into Eq.~\eqref{eq:Jneq} and using $G^A(\bR,\varepsilon)=\big[G^R(-\bR,\varepsilon)\big]^\dagger$, the two mixed products $G^R(\bR,\varepsilon)\,G^A(-\bR,\varepsilon)$ cancel, and we obtain the equilibrium RKKY coupling
\begin{equation}
\begin{aligned}
J^{(0)}_{ab}(\bR)&=-\frac{\Jc^2}{\pi}\,{\rm Im}\int_{-\infty}^{\eF}\!d\varepsilon\;
{\rm Tr}\Big[\sigma_aG^R(\bR,\varepsilon)\,\sigma_bG^R(-\bR,\varepsilon)\Big].
\end{aligned}
\label{eq:J0GF}
\end{equation}

\textit{Retarded GF in real space:}
In a homogeneous steady state every function depends on $\mathbf r_1-\mathbf r_2$ and
$t_1-t_2$ only, and the GF is,
\begin{equation}
G(\bR,\epsilon)=\int\!\frac{d^2k}{(2\pi)^2}\;
e^{i\bk\cdot\bR}\,G(\bk,\varepsilon) .
\label{eq:FTconvo}
\end{equation}
We rescale the momenta for each spin as $k_j=\sqrt{m_{j\sigma}}\,p_{\sigma,j}$ ($j=x,y$), with $\mathbf p_\sigma=(p_{\sigma,x},p_{\sigma,y})$ and $p_\sigma=|\mathbf p_\sigma|$, so that $d^2k=M\,d^2p_\sigma$ with $M=\sqrt{m_{x\sigma}m_{y\sigma}}$, the band energy becomes isotropic, $\varepsilon_\sigma(\bk)=\sigma\Jp+\hbar^2p_\sigma^2/2$, and $\bk\cdot\bR=\mathbf p_\sigma\cdot\bm\rho_\sigma$ with $\bm\rho_\sigma=(\sqrt{m_{x\sigma}}R_x,\sqrt{m_{y\sigma}}R_y)$. Then
\begin{equation}
G^R_\sigma(\bR,\varepsilon)=\frac{2M}{\hbar^2}\int\!\frac{d^2p_\sigma}{(2\pi)^2}\,
\frac{e^{i\mathbf p_\sigma\cdot\bm\rho_\sigma}}{\kappa_\sigma^2-p_\sigma^2+i0},
\label{eq:grelliptic}
\end{equation}
which depends on $\bR$ only through $\kappa_\sigma\rho_\sigma\equiv k_\sigma(\phi,\varepsilon)R$, where $\kappa_\sigma=\sqrt{2(\varepsilon-\sigma\Jp)}/\hbar$, $\rho_\sigma=R\sqrt{m_\sigma(\phi)}$, $m_\sigma(\phi)=m_{x\sigma}\cos^2\phi+m_{y\sigma}\sin^2\phi$, and $k_\sigma(\phi,\varepsilon)=\sqrt{2m_\sigma(\phi)(\varepsilon-\sigma\Jp)}/\hbar$ is the wave number along the bond direction $\phi$.
Performing the angular integral in polar coordinates about $\bm\rho_\sigma$ with $\int_0^{2\pi}d\vartheta_\sigma\,e^{ip_\sigma\rho_\sigma\cos\vartheta_\sigma}=2\pi J_0(p_\sigma\rho_\sigma)$, we obtain $G^R_\sigma(\bR,\varepsilon)=(M/\pi\hbar^2)\,\mathcal I_\sigma$, where $\mathcal I_\sigma=\int_0^\infty dp_\sigma\,p_\sigma J_0(p_\sigma\rho_\sigma)/(\kappa_\sigma^2-p_\sigma^2+i0)$. After some algebra (see SI) and performing the contour integration we obtain, $\mathcal I_\sigma=-\tfrac{i\pi}{2}\,H_0^{(1)}(\kappa_\sigma\rho_\sigma)$ and the GF becomes,
\begin{equation}
G^R_\sigma(\bR,\varepsilon)=-\frac{iM}{2\hbar^2}\,H_0^{(1)}\big(k_\sigma(\phi,\varepsilon)R\big).
\label{eq:G}
\end{equation}
It depends on $\bR$ only through $R$ and $m_\sigma(\phi)$, so
$G^R_\sigma(-\bR,\varepsilon)=G^R_\sigma(\bR,\varepsilon)$, which
follows from $m_\sigma(\phi+\pi)=m_\sigma(\phi)$
and as a consequence the trace in Eq.~\eqref{eq:J0GF},
\begin{equation}
{\rm Tr}\big[\sigma_a\,G^R(\bR,\varepsilon)\,\sigma_b\,G^R(-\bR,\varepsilon)\big]
=\sum_{\sigma\sigma'}c^{ab}_{\sigma\sigma'}\,G^R_{\sigma'}(\bR,\varepsilon)\,G^R_\sigma(-\bR,\varepsilon).
\end{equation}
The coefficients $c^{ab}_{\sigma\sigma'}$ are listed in Table~\ref{tab:cab} (see SI for detailed calculation). 
\begin{table}
\centering
\caption{Spin-channel coefficients $c^{ab}_{\sigma\sigma'}=\sigma^a_{\sigma\sigma'}\sigma^b_{\sigma'\sigma}$.}
\label{tab:cab}
\begin{tabular}{lcccc}
\toprule
$(a,b)$ & $c_{\up\up}$ & $c_{\dn\dn}$ & $c_{\up\dn}$ & $c_{\dn\up}$\\
\midrule
$(x,x)$ & 0 & 0 & $1$ & $1$\\
$(y,y)$ & 0 & 0 & $1$ & $1$\\
$(z,z)$ & 1 & 1 & 0 & 0\\
$(x,y)$ & 0 & 0 & $i$ & $-i$\\
$(y,x)$ & 0 & 0 & $-i$ & $i$\\
$(x,z)$, $(z,x)$, $(y,z)$, $(z,y)$ & 0 & 0 & 0 & 0\\
\bottomrule
\end{tabular}
\end{table}
 With Table~\ref{tab:cab},
 \begin{equation*}
\begin{array}{rcl}
{\rm Tr}\big[\sigma_x\,G^R(\bR,\varepsilon)\,\sigma_x\,G^R(-\bR,\varepsilon)\big]&=&2\,G^R_\up(\bR,\varepsilon)\,G^R_\dn(\bR,\varepsilon),\\
{\rm Tr}\big[\sigma_z\,G^R(\bR,\varepsilon)\,\sigma_z\,G^R(-\bR,\varepsilon)\big]&=&\big[G^R_\up(\bR,\varepsilon)\big]^2+\big[G^R_\dn(\bR,\varepsilon)\big]^2,\\
{\rm Tr}\big[\sigma_x\,G^R(\bR,\varepsilon)\,\sigma_y\,G^R(-\bR,\varepsilon)\big]&=&i\big[G^R_\dn(\bR,\varepsilon)\,G^R_\up(\bR,\varepsilon)-G^R_\up(\bR,\varepsilon)\,G^R_\dn(\bR,\varepsilon)\big]=0 .
\end{array}
\end{equation*}
Consequently, the equilibrium RKKY couplings become,
\begin{equation}
\begin{array}{rcl}
J_\perp^{(0)}(\bR)\equiv J^{(0)}_{xx}(\bR)=J^{(0)}_{yy}(\bR)&=&-\dfrac{2\Jc^2}{\pi}\,{\rm Im}\displaystyle\int_{-\infty}^{\eF}\!d\varepsilon\;G^R_\up(\bR,\varepsilon)\,G^R_\dn(\bR,\varepsilon),\\
J_{zz}^{(0)}(\bR)&=&-\dfrac{\Jc^2}{\pi}\,{\rm Im}\displaystyle\int_{-\infty}^{\eF}\!d\varepsilon\;\Big\{\big[G^R_\up(\bR,\varepsilon)\big]^2+\big[G^R_\dn(\bR,\varepsilon)\big]^2\Big\},\\
D_z^{(0)}(\bR)&=&-J^{(0)}_{xy}(\bR)=0 .
\end{array}
\label{eq:J0comp}
\end{equation}
Using $G^R_\sigma(\bR,\varepsilon)\,G^R_{\sigma'}(\bR,\varepsilon)=-\frac{M^2}{4\hbar^4}\,H_0^{(1)}\big(k_\sigma(\phi,\varepsilon)R\big)\,H_0^{(1)}\big(k_{\sigma'}(\phi,\varepsilon)R\big)$ we obtain,
\begin{equation}
\begin{array}{rcl}
J_\perp^{(0)}(\bR)&=&\dfrac{\Jc^2M^2}{2\pi\hbar^4}\,{\rm Im}\displaystyle\int_{-\infty}^{\eF}\!d\varepsilon\;
H_0^{(1)}\big(k_\up(\phi,\varepsilon)R\big)\,H_0^{(1)}\big(k_\dn(\phi,\varepsilon)R\big),\\
J_{zz}^{(0)}(\bR)&=&\dfrac{\Jc^2 M^2}{4\pi\hbar^4}\displaystyle\sum_\sigma \,{\rm Im}\int_{-\infty}^{\eF}\!d\varepsilon\;
\Big[H_0^{(1)}\big(k_\sigma(\phi,\varepsilon)R\big)\Big]^2 .
\end{array}
\label{eq:J0H}
\end{equation}
We assume that the Fermi energy lies well above both band bottoms, $\eF>|\Jp|$, so that both
spin bands are occupied and $k^F_\sigma(\phi)\equiv k_\sigma(\phi,\eF)$ is real for both spins.
After the energy integration (see Sec.~\sref{s-zeroth-order-RKKY} of the SI),
\begin{align}
J_{zz}^{(0)}(\bR)&=\frac{\Jc^2M^2}{2\pi\hbar^4}\sum_\sigma \,(\eF-\sigma\Jp)\,
\Big[J_0\big(k^F_\sigma(\phi)R\big)Y_0\big(k^F_\sigma(\phi)R\big)+J_1\big(k^F_\sigma(\phi)R\big)Y_1\big(k^F_\sigma(\phi)R\big)\Big],
\label{eq:Jzz0exact}\\
J_\perp^{(0)}(\bR)&=\frac{\Jc^2M^2}{2\pi\hbar^4}\,{\rm Im}\int_{-|\Jp|}^{\eF}\!d\varepsilon\;
H_0^{(1)}\big(k_\up(\phi,\varepsilon)R\big)\,H_0^{(1)}\big(k_\dn(\phi,\varepsilon)R\big).
\label{eq:Jperp0general}
\end{align}
The Eq.~\eqref{eq:Jperp0general} can not be evaluated analytically and we therefore evaluate numerically.

\textit{Asymptotic form}: For $k_\sigma(\phi,\varepsilon)R\gg1$ we use the large-argument forms
\begin{equation}
\begin{array}{rcl}
H_0^{(1)}\big(k_\sigma(\phi,\varepsilon)R\big)&\simeq&\sqrt{\dfrac{2}{\pi k_\sigma(\phi,\varepsilon)R}}\;
e^{i\left[k_\sigma(\phi,\varepsilon)R-\pi/4\right]},\\
J_\nu\big(k^F_\sigma(\phi)R\big)&\simeq&\sqrt{\dfrac{2}{\pi k^F_\sigma(\phi)R}}
\left[\cos\Big(k^F_\sigma(\phi)R-\dfrac{\nu\pi}{2}-\dfrac{\pi}{4}\Big)
-\dfrac{4\nu^2-1}{8k^F_\sigma(\phi)R}\,\sin\Big(k^F_\sigma(\phi)R-\dfrac{\nu\pi}{2}-\dfrac{\pi}{4}\Big)\right],\\
Y_\nu\big(k^F_\sigma(\phi)R\big)&\simeq&\sqrt{\dfrac{2}{\pi k^F_\sigma(\phi)R}}
\left[\sin\Big(k^F_\sigma(\phi)R-\dfrac{\nu\pi}{2}-\dfrac{\pi}{4}\Big)
+\dfrac{4\nu^2-1}{8k^F_\sigma(\phi)R}\,\cos\Big(k^F_\sigma(\phi)R-\dfrac{\nu\pi}{2}-\dfrac{\pi}{4}\Big)\right],
\end{array}
\label{eq:Besselasym}
\end{equation}
where $J_\nu$ and $Y_\nu$ are kept to next-to-leading order in $1/k^F_\sigma(\phi)R$. In
Eq.~\eqref{eq:Jzz0exact} the leading terms of $J_0Y_0$ and $J_1Y_1$ cancel, and the
next-to-leading terms give
\begin{equation}
J_0\big(k^F_\sigma(\phi)R\big)Y_0\big(k^F_\sigma(\phi)R\big)+J_1\big(k^F_\sigma(\phi)R\big)Y_1\big(k^F_\sigma(\phi)R\big)
\simeq-\frac{\sin\big(2k^F_\sigma(\phi)R\big)}{\pi\,[k^F_\sigma(\phi)R]^2}.
\end{equation}
With $\eF-\sigma\Jp=\hbar^2[k^F_\sigma(\phi)]^2/2m_\sigma(\phi)$ and
$v^F_\sigma(\phi)=\hbar k^F_\sigma(\phi)/m_\sigma(\phi)$,
\begin{equation}
J_{zz}^{(0)}(\bR)\simeq-\frac{\Jc^2M^2}{4\pi^2\hbar^3}\sum_\sigma
\frac{v^F_\sigma(\phi)}{k^F_\sigma(\phi)}\,\frac{\sin\big(2k^F_\sigma(\phi)R\big)}{R^2}.
\label{eq:Jzz0asym}
\end{equation}

\section{1st-order in {\bf E}: Non-equillibrium RKKY  }
\label{sec: non-equillibrium-RKKY}
Using $A_\sigma(\bk,\varepsilon)=2\pi\,\delta\big(\varepsilon-\varepsilon_\sigma(\bk)\big)$ and $\nabla_{\bk}\,\theta\big(\varepsilon-\varepsilon_\sigma(\bk)\big)=-\nabla_{\bk}\varepsilon_\sigma(\bk)\,\delta\big(\varepsilon-\varepsilon_\sigma(\bk)\big)$, the momentum-space lesser correction, Eq.~\eqref{eq:gklesser}, becomes
 $\delta G^<_\sigma(\bk,\varepsilon)
=-2\pi i\,\big[-f_0'(\varepsilon)\big]\,\dk\cdot\nabla_{\bk}\,\theta\big(\varepsilon-\varepsilon_\sigma(\bk)\big)$,
and the result is valid for an arbitrary band $\varepsilon_\sigma(\bk)$.
Using  $\nabla_{\bk}e^{i\bk\cdot\bR}=i\bR\,e^{i\bk\cdot\bR}$ and $\int d^2k\,\dk\cdot\nabla_{\bk}\big(e^{i\bk\cdot\bR}\theta\big)=\oint_{|\bk|\to\infty}dS_{\bk}\,(\dk\cdot\hat{\mathbf n})\,e^{i\bk\cdot\bR}\,\theta\big(\varepsilon-\varepsilon_\sigma(\bk)\big)=0$, since $\theta=0$ outside the Fermi contour,  the FT of the correction to the lesser GF above is,
\begin{equation}
\begin{array}{rcl}
\delta G^<_\sigma(\bR,\varepsilon)
&=&-2\pi i\,\big[-f_0'(\varepsilon)\big]\displaystyle\int\!\frac{d^2k}{(2\pi)^2}\,e^{i\bk\cdot\bR}\,\dk\cdot\nabla_{\bk}\,\theta\big(\varepsilon-\varepsilon_\sigma(\bk)\big)\\
&=&\big[-f_0'(\varepsilon)\big]a_\sigma(\bR,\varepsilon),
\end{array}
\label{eq:delg_intermsof_a}
\end{equation}
where,  $a_\sigma(\bR,\varepsilon)=-2\pi\,(\dk\cdot\bR)\,N_\sigma(\bR,\varepsilon)$ with,
\begin{equation}
\begin{array}{rcl}
N_\sigma(\bR,\varepsilon)&\equiv&\displaystyle\int\!\frac{d^2k}{(2\pi)^2}\,e^{i\bk\cdot\bR}\,\theta\big(\varepsilon-\varepsilon_\sigma(\bk)\big).
\end{array}
\label{eq:Ndef}
\end{equation}
For even bands, $\theta(\varepsilon-\varepsilon_\sigma(\bk))$ is real and even in $\bk$, so $N_\sigma(\bR,\varepsilon)$ is real and even in $\bR$; since $\dk\cdot\bR$ is real and odd,
\begin{equation}
a_\sigma(\bR,\varepsilon)\in\mathbb R,\qquad a_\sigma(-\bR,\varepsilon)=-a_\sigma(\bR,\varepsilon).
\label{eq:aprops}
\end{equation}
The whole effect of the current sits in the factor $\dk\cdot\bR$: shifting the occupied region by $\dk$ multiplies its real-space transform by a phase, $\int d^2k\,(2\pi)^{-2}e^{i\bk\cdot\bR}\,\theta\big(\varepsilon-\varepsilon_\sigma(\bk-\dk)\big)
=e^{i\dk\cdot\bR}N_\sigma(\bR,\varepsilon)\simeq(1+i\,\dk\cdot\bR)\,N_\sigma(\bR,\varepsilon)$.
To calculate $N_\sigma(\bR,\varepsilon)$, we use the same rescaling $k_j=\sqrt{m_{j\sigma}}\,p_{\sigma,j}$ that maps the elliptic band onto a circular one in Eq.~\eqref{eq:grelliptic} and obtain (see SI),
\begin{align}
N_\sigma(\bR,\varepsilon)&=M\int\!\frac{d^2p_\sigma}{(2\pi)^2}\,e^{i\mathbf p_\sigma\cdot\bm\rho_\sigma}\,\theta(\kappa_\sigma-p_\sigma)=\frac{M\kappa_\sigma}{2\pi\rho_\sigma}J_1(\kappa_\sigma\rho_\sigma),\\
a_\sigma(\bR,\varepsilon)&=-2\pi\,(\dk\cdot\bR)\,\dfrac{M\,\kappa_\sigma}{2\pi\rho_\sigma}\,J_1(\kappa_\sigma\rho_\sigma).
\label{eq:asigmaRepsilon}
\end{align}
As a consequence, from Eq.~\eqref{eq:delg_intermsof_a},
$\delta G^<_\sigma(\bR,\varepsilon)
=-2\pi\big[-f_0'(\varepsilon)\big]\,(\dk\cdot\bR)\,({M\,\kappa_\sigma}/{2\pi\rho_\sigma})\,J_1(\kappa_\sigma\rho_\sigma)$.
Writing $\delta G^<(\bR,\varepsilon)=\big[-f_0'(\varepsilon)\big]\,\hat a(\bR,\varepsilon)$ with
$\hat a(\bR,\varepsilon)={\rm diag}\big(a_\up(\bR,\varepsilon),a_\dn(\bR,\varepsilon)\big)$ in spin space, the current-induced part of Eq.~\eqref{eq:Jsplit} reads
\begin{equation}
\begin{aligned}
\delta J_{ab}(\bR)=-\frac{i\Jc^2}{2\pi}\int d\varepsilon\,\big[-f_0'(\varepsilon)\big]
\Big\{&{\rm Tr}\big[\sigma_a\,G^R(\bR,\varepsilon)\,\sigma_b\,\hat a(-\bR,\varepsilon)\big]\\
+\,&{\rm Tr}\big[\sigma_a\,\hat a(\bR,\varepsilon)\,\sigma_b\,G^A(-\bR,\varepsilon)\big]\Big\}.
\end{aligned}
\label{eq:dJGF}
\end{equation}
Since $a_\sigma(-\bR,\varepsilon)=-a_\sigma(\bR,\varepsilon)$ (Eq.~\eqref{eq:aprops}), the first trace equals $-\mathcal T_{ab}(\bR,\varepsilon)$, with
$\mathcal T_{ab}(\bR,\varepsilon)\equiv{\rm Tr}\big[\sigma_a\,G^R(\bR,\varepsilon)\,\sigma_b\,\hat a(\bR,\varepsilon)\big]$.
Since $a_\sigma(\bR,\varepsilon)$ is real, $G^A(-\bR,\varepsilon)=\big[G^R(\bR,\varepsilon)\big]^\dagger$, the Pauli matrices are Hermitian and the trace is cyclic,
the second trace equals ${\rm Tr}\big[\big(\sigma_a\,G^R(\bR,\varepsilon)\,\sigma_b\,\hat a(\bR,\varepsilon)\big)^\dagger\big]=\mathcal T_{ab}^*(\bR,\varepsilon)$.
The braces therefore give $-\big[\mathcal T_{ab}(\bR,\varepsilon)-\mathcal T_{ab}^*(\bR,\varepsilon)\big]=-2i\,{\rm Im}\,\mathcal T_{ab}(\bR,\varepsilon)$, and
\begin{equation}
\delta J_{ab}(\bR)=-\frac{\Jc^2}{\pi}\int d\varepsilon\,\big[-f_0'(\varepsilon)\big]\,
{\rm Im}\,{\rm Tr}\big[\sigma_a\,G^R(\bR,\varepsilon)\,\sigma_b\,\hat a(\bR,\varepsilon)\big].
\label{eq:dJIm}
\end{equation}
At $T=0$, $-f_0'(\varepsilon)=\delta(\varepsilon-\eF)$, and the energy integral collapses onto the
FS and we obtain,
\begin{equation}
\delta J_{ab}(\bR)=-\frac{\Jc^2}{\pi}\,
{\rm Im}\,{\rm Tr}\big[\sigma_a\,G^R(\bR,\eF)\,\sigma_b\,\hat a(\bR, \eF)\big].
\label{eq:dJFS}
\end{equation}
In \eqref{eq:G} using the relationship for an arbitrary $x$,   $H_0^{(1)}(x)=J_0(x)+iY_0(x)$ and as
defined earlier $k^F_\sigma(\phi)\equiv k_\sigma(\phi,\eF)$, we write,
\begin{equation}
\begin{array}{rcl}
G^R_\sigma(\bR,\eF)&=&g_\sigma(\bR,\eF)=g'_\sigma(\bR,\eF)+i\,g''_\sigma(\bR,\eF),\\
g'_\sigma(\bR,\eF)&=&\dfrac{M}{2\hbar^2}\,Y_0\big(k^F_\sigma(\phi)R\big),\\
g''_\sigma(\bR,\eF)&=&-\dfrac{M}{2\hbar^2}\,J_0\big(k^F_\sigma(\phi)R\big).
\end{array}
\label{eq:gparts}
\end{equation}
After some simple algebra,
\begin{equation}
{\rm Tr}\big[\sigma_a\,G^R(\bR,\eF)\,\sigma_b\,\hat a(\bR,\eF)\big]=\sum_{\sigma\sigma'}c^{ab}_{\sigma\sigma'}\,g_{\sigma'}(\bR,\eF)\,a_\sigma(\bR,\eF),
\end{equation}
where $c^{ab}_{\sigma\sigma'}=\sigma^a_{\sigma\sigma'}\,\sigma^b_{\sigma'\sigma}$ are given in Table~\ref{tab:cab} and the only terms survive are, 
\begin{equation}
\begin{array}{rcl}
{\rm Tr}\big[\sigma_x\,G^R(\bR,\eF)\,\sigma_x\,\hat a(\bR,\eF)\big]&=&g_\dn(\bR,\eF)\,a_\up(\bR,\eF)+g_\up(\bR,\eF)\,a_\dn(\bR,\eF),\\
{\rm Tr}\big[\sigma_z\,G^R(\bR,\eF)\,\sigma_z\,\hat a(\bR,\eF)\big]&=&g_\up(\bR,\eF)\,a_\up(\bR,\eF)+g_\dn(\bR,\eF)\,a_\dn(\bR,\eF),\\
{\rm Tr}\big[\sigma_x\,G^R(\bR,\eF)\,\sigma_y\,\hat a(\bR,\eF)\big]&=&i\,\big[g_\dn(\bR,\eF)\,a_\up(\bR,\eF)-g_\up(\bR,\eF)\,a_\dn(\bR,\eF)\big].
\end{array}
\end{equation}
Since $a_\sigma(\bR,\eF)$ is real, ${\rm Im}\big[g_\sigma(\bR,\eF)\,a_{\sigma'}(\bR,\eF)\big]=g''_\sigma(\bR,\eF)\,a_{\sigma'}(\bR,\eF)$ and ${\rm Im}\big[i\,g_\sigma(\bR,\eF)\,a_{\sigma'}(\bR,\eF)\big]=g'_\sigma(\bR,\eF)\,a_{\sigma'}(\bR,\eF)$.
Equation~\eqref{eq:dJFS} then gives
\begin{equation}
\begin{array}{rcl}
\delta J_\perp(\bR)\equiv\delta J_{xx}(\bR)=\delta J_{yy}(\bR)&=&-\dfrac{\Jc^2}{\pi}\big[a_\up(\bR,\eF)\,g''_\dn(\bR,\eF)+a_\dn(\bR,\eF)\,g''_\up(\bR,\eF)\big],\\
\delta J_{zz}(\bR)&=&-\dfrac{\Jc^2}{\pi}\big[a_\up(\bR,\eF)\,g''_\up(\bR,\eF)+a_\dn(\bR,\eF)\,g''_\dn(\bR,\eF)\big],\\
D_z^{(1)}(\bR)=-\delta J_{xy}(\bR)=\delta J_{yx}(\bR)&=&\dfrac{\Jc^2}{\pi}\big[a_\up(\bR,\eF)\,g'_\dn(\bR,\eF)-a_\dn(\bR,\eF)\,g'_\up(\bR,\eF)\big].
\end{array}
\label{eq:dJcomp}
\end{equation}
From  Eq.~\eqref{eq:asigmaRepsilon}  and Eq.~ \eqref{eq:gparts} and using $v^F_\sigma(\phi)=({\hbar\,k^F_\sigma(\phi)})/({m_\sigma(\phi)})$ we obtain,
\begin{align}
\delta J_{zz}(\bR)&=-\frac{\Jc^2M^2}{2\pi\hbar^3}\,(\dk\cdot\Rh)\sum_\sigma \,v^F_\sigma(\phi)\,
J_1\big(k^F_\sigma(\phi)R\big)\,J_0\big(k^F_\sigma(\phi)R\big),
\label{eq:Jzznr}\\
\delta J_\perp(\bR)&=-\frac{\Jc^2M^2}{2\pi\hbar^3}\,(\dk\cdot\Rh)
\Big[v^F_\up(\phi)\,J_1\big(k^F_\up(\phi)R\big)\,J_0\big(k^F_\dn(\phi)R\big)\nonumber\\
&\hspace{9em}+v^F_\dn(\phi)\,J_1\big(k^F_\dn(\phi)R\big)\,J_0\big(k^F_\up(\phi)R\big)\Big],
\label{eq:Jperpnr}\\
\delta D_z(\bR)&=\frac{\Jc^2M^2}{2\pi\hbar^3}\,(\dk\cdot\Rh)
\Big[v^F_\dn(\phi)\,J_1\big(k^F_\dn(\phi)R\big)\,Y_0\big(k^F_\up(\phi)R\big)\nonumber\\
&\hspace{9em}-v^F_\up(\phi)\,J_1\big(k^F_\up(\phi)R\big)\,Y_0\big(k^F_\dn(\phi)R\big)\Big].
\label{eq:Dexact}
\end{align}
\paragraph{Parity under $\bR\to-\bR$.} All current-induced couplings in
Eqs.~\eqref{eq:Jzznr}--\eqref{eq:Dexact} are proportional to $\dk\cdot\Rh$, which is odd under
$\bR\to-\bR$, whereas $k^F_\sigma(\phi)$, $v^F_\sigma(\phi)$, and hence the Bessel-function
factors are even. Therefore $\delta J_{ab}(-\bR)=-\delta J_{ab}(\bR)$. For the antisymmetric
component this gives $\delta J_{xy}(-\bR)=-\delta J_{xy}(\bR)=\delta J_{yx}(\bR)$, so the
condition $J_{ab}(\bR)=J_{ba}(-\bR)$, obeyed by any coupling derived from a spin Hamiltonian, is
satisfied, and $\delta D_z$ is a genuine Dzyaloshinskii--Moriya interaction. The symmetric
components instead obey $\delta J_{aa}(-\bR)=-\delta J_{aa}(\bR)$, so $\delta J_\perp$ and
$\delta J_{zz}$ violate this condition: no spin energy $E(\bS_1,\bS_2)$ yields them, and they
enter only the equations of motion as nonconservative couplings.

Physically, these symmetric terms act on the two spins asymmetrically. Let
$\delta J^{\rm s}={\rm diag}(\delta J_\perp,\delta J_\perp,\delta J_{zz})$ and let $\bR$ point
from $\bS_1$ to $\bS_2$. The exchange field exerted on $\bS_2$ by $\bS_1$ is then
$(J^{(0)}+\delta J^{\rm s})\bS_1$, whereas that exerted on $\bS_1$ by $\bS_2$ is
$(J^{(0)}-\delta J^{\rm s})\bS_2$. The two mutual fields thus differ by $2\,\delta J^{\rm s}$,
with a sign set by the direction of the current relative to the bond.
\subsection{Long-distance forms}

For $k^F_\sigma(\phi)R\gg1$, $J_1$ and $Y_0$ coincide at leading order
[Eq.~\eqref{eq:Besselasym}], so the two products in Eq.~\eqref{eq:Dexact} share one
trigonometric factor. With
$K(\phi)=k^F_\up(\phi)+k^F_\dn(\phi)$ and $\Delta k(\phi)=k^F_\up(\phi)-k^F_\dn(\phi)$,
\begin{align}
\delta D_z(\bR)&\simeq\frac{\Jc^2M^2\,\big[v^F_\dn(\phi)-v^F_\up(\phi)\big]}{2\pi^2\hbar^3\sqrt{k^F_\up(\phi)k^F_\dn(\phi)}}\;(\dk\cdot\Rh)\;
\frac{\cos\big(\Delta k(\phi)R\big)-\sin\big(K(\phi)R\big)}{R},
\label{eq:Dasym}\\
\delta J_\perp(\bR)&\simeq\frac{\Jc^2M^2}{2\pi^2\hbar^3\sqrt{k^F_\up(\phi)k^F_\dn(\phi)}}\;(\dk\cdot\Rh)\nonumber\\
&\quad\times\frac{\big[v^F_\up(\phi)+v^F_\dn(\phi)\big]\cos\big(K(\phi)R\big)+\big[v^F_\dn(\phi)-v^F_\up(\phi)\big]\sin\big(\Delta k(\phi)R\big)}{R},
\label{eq:Jperpasym}\\
\delta J_{zz}(\bR)&\simeq\frac{\Jc^2M^2}{2\pi^2\hbar^3}\;(\dk\cdot\Rh)\sum_\sigma\frac{v^F_\sigma(\phi)}{k^F_\sigma(\phi)}\,
\frac{\cos\big(2k^F_\sigma(\phi)R\big)}{R}.
\label{eq:Jzzasym}
\end{align}
For comparison, in the same limit $k^F_\sigma(\phi)R\gg1$ the equilibrium couplings read
\begin{equation}
\begin{array}{rcl}
J_\perp^{(0)}(\bR)&\simeq&-\dfrac{\Jc^2M^2\,\bar v(\phi)}{2\pi^2\hbar^3\sqrt{k^F_\up(\phi)k^F_\dn(\phi)}}\,
\dfrac{\sin\big(K(\phi)R\big)}{R^2},\qquad
\bar v(\phi)=\dfrac{2v^F_\up(\phi)v^F_\dn(\phi)}{v^F_\up(\phi)+v^F_\dn(\phi)},\\
J_{zz}^{(0)}(\bR)&\simeq&-\dfrac{\Jc^2M^2}{4\pi^2\hbar^3}\displaystyle\sum_\sigma\frac{v^F_\sigma(\phi)}{k^F_\sigma(\phi)}\,
\frac{\sin\big(2k^F_\sigma(\phi)R\big)}{R^2}.
\end{array}
\label{eq:J0asym}
\end{equation}
Comparing Eqs.~\eqref{eq:Dasym}--\eqref{eq:Jzzasym} with Eq.~\eqref{eq:J0asym}, the
current-induced couplings decay as $R^{-1}$, one power slower than the equilibrium couplings,
which decay as $R^{-2}$. Both sets oscillate with the fast wave numbers $K(\phi)$ and
$2k^F_\sigma(\phi)$, but only $\delta D_z$ and $\delta J_\perp$ contain the slow beat terms
$\cos\big(\Delta k(\phi)R\big)$ and $\sin\big(\Delta k(\phi)R\big)$, which have no counterpart in
$J^{(0)}_{ab}$.

\subsection{Comparison of the coupling strengths}

To compare the strengths of the different couplings, we evaluate the long-distance forms,
Eqs.~\eqref{eq:Dasym}--\eqref{eq:J0asym}, for a bond along the current, $\phi=0$ with
$\dk\parallel\hat x$, i.e., $\dk\cdot\Rh=\delta k$, and replace each oscillatory factor by its
envelope. We use the reference values $k_F=\sqrt{2m\eF}/\hbar$ and $v_F=\hbar k_F/m$ of the
unsplit band. For the $zz$ channel the band sum is exact,
$\sum_\sigma v^F_\sigma(0)/k^F_\sigma(0)=\sum_\sigma\hbar/m_{x\sigma}=2\hbar/m$. For the other
channels we keep the leading order in $b$ and $\Jp/\eF$, where
$v^F_\sigma(0)/k^F_\sigma(0)\simeq\hbar/m$, $\sqrt{k^F_\up(0)k^F_\dn(0)}\simeq k_F$,
$v^F_\dn(0)-v^F_\up(0)\simeq v_F\big(\Jp/\eF-b\big)$, and
$\Delta k(0)\simeq-k_F\big(\Jp/\eF+b\big)$.
The amplitudes are then
\begin{equation}
\begin{array}{rclrcl}
\big|J_{zz}^{(0)}\big|&\sim&\dfrac{\Jc^2M^2}{2\pi^2\hbar^2m}\,\dfrac{1}{R^2},\qquad&
\big|\delta J_{zz}\big|&\sim&\dfrac{\Jc^2M^2}{\pi^2\hbar^2m}\,\dfrac{\delta k}{R},\\
\big|J_\perp^{(0)}\big|&\sim&\dfrac{\Jc^2M^2}{2\pi^2\hbar^2m}\,\dfrac{1}{R^2},\qquad&
\big|\delta J_\perp\big|&\sim&\dfrac{\Jc^2M^2}{\pi^2\hbar^2m}\,\dfrac{\delta k}{R},\\
&&&\big|\delta D_z\big|&\sim&\dfrac{\Jc^2M^2}{2\pi^2\hbar^2m}\,\Big|\dfrac{\Jp}{\eF}-b\Big|\,\dfrac{\delta k}{R}.
\end{array}
\label{eq:amplitudes}
\end{equation}
For $\delta D_z$ we kept only the slow term and set $\cos(\Delta k\,R)\to1$; this is the
envelope of the sign-definite part, and it equals the actual value only for
$R\ll1/|\Delta k|$. For $\delta J_\perp$ we dropped the beat term, which is smaller by
$|v^F_\dn-v^F_\up|/(v^F_\up+v^F_\dn)\sim\frac12|\Jp/\eF-b|$. The ratios follow as
\begin{equation}
\frac{|\delta J_{zz}|}{|J_{zz}^{(0)}|}\sim\frac{|\delta J_\perp|}{|J_\perp^{(0)}|}\sim2\,\delta k\,R,\qquad
\frac{|\delta D_z|}{|J_\perp^{(0)}|}\sim\Big|\frac{\Jp}{\eF}-b\Big|\,\delta k\,R,\qquad
\frac{|\delta D_z|}{|\delta J_{zz}|}\sim\frac12\Big|\frac{\Jp}{\eF}-b\Big|.
\label{eq:ratios}
\end{equation}
All current-induced couplings grow relative to the equilibrium ones linearly in $\delta k\,R$.
The symmetric corrections are insensitive to the spin splitting, whereas the
Dzyaloshinskii--Moriya term is further reduced by the splitting factor $|\Jp/\eF-b|$, i.e.\ by
$|b|$ for the pure altermagnet and by $\Jp/\eF$ for the ferromagnet.

For a numerical estimate we take a bond along the current with $k_FR=20$ and
$|\Jp/\eF-b|=0.2$; the ratios then scale with $v_d/v_F=\delta k/k_F$. For a pure altermagnet
with $b=0.2$ this gives $|\Delta k|R\approx4$, so the values quoted for $\delta D_z$ are
envelope values. A metallic altermagnetic film ($n\approx10^{28}\,{\rm m^{-3}}$,
$j=10^{11}\,{\rm A/m^2}$, free-electron $v_F\approx8\times10^{5}$\,m/s) has
$v_d=j/ne\approx60$\,m/s and $v_d/v_F\approx8\times10^{-5}$, which gives
$|\delta J|/|J^{(0)}|\approx3\times10^{-3}$ and $|\delta D_z|/|J^{(0)}|\approx3\times10^{-4}$.
A low-density 2D layer ($n\approx10^{16}\,{\rm m^{-2}}$, $m^*\approx0.1\,m_e$, sheet current
density $j_s=5$\,A/m) gives $v_d/v_F\approx10^{-2}$, $|\delta J|/|J^{(0)}|\approx0.4$ and
$|\delta D_z|/|J^{(0)}|\approx0.04$. Here $\delta k\,R\approx0.2$, which is at the edge of the
linear-response regime.

\section{Numerical illustration of the results}
\label{sec:figures}

 In this section we evaluate the RKKY couplings numerically, using Eqs.~\eqref{eq:Jzznr}--\eqref{eq:Dexact} for the current-induced
couplings and Eq.~\eqref{eq:Jzz0exact} for $J_{zz}^{(0)}$, and integrating
Eq.~\eqref{eq:Jperp0general} numerically for $J_\perp^{(0)}$.
 We set $\hbar=m=k_F=1$, so that $\eF=1/2$ and $v_F=1$, and measure lengths in
units of $1/k_F$. 
Couplings are given in units of $\Jc^2mk_F^2/\hbar^2$; the current-induced
couplings are linear in the drift and are shown per unit $\delta k/k_F$. 
We specify directions by the drift $\dk=-e\tau\mathbf E/\hbar$, which for electrons is antiparallel to the current $\mathbf j$. Unless stated
otherwise, we take $b=0.10$ for the altermagnet, $\Jp/\eF=0.10$ for the ferromagnet, and $b=0.10$,
$\Jp/\eF=0.05$ for the mixed case.

\begin{figure}
\centering
\subfloat[]{\includegraphics[width=0.40\textwidth]{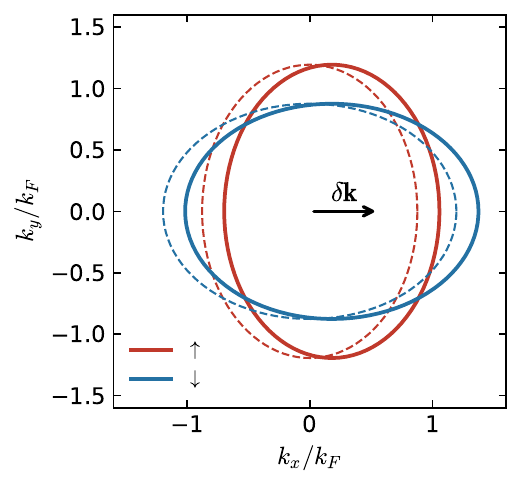}}\hspace{0.05\textwidth}
\subfloat[]{\includegraphics[width=0.42\textwidth]{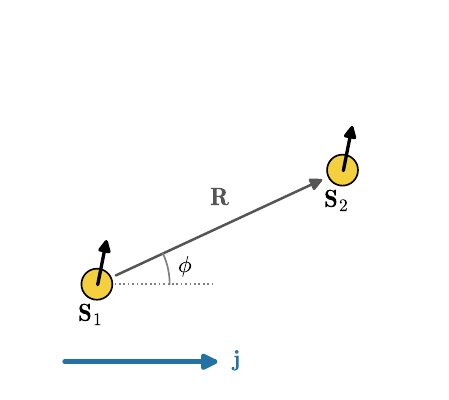}}
\caption{Schematic of the RKKY mechanism. (a)~Fermi contours of the two spin bands of a $d$-wave
altermagnet in equilibrium (dashed) and in the current-carrying state (solid), rigidly displaced
by the drift $\dk$. 
(b)~Geometry: impurity spins $\bS_1$ and $\bS_2$ in a two-dimensional altermagnet with exchange
splitting $\Jp$, separated by $\bR$ at angle $\phi$ to the $x$ axis, and the current
$\mathbf j$.}
\label{fig:mechanism}
\end{figure}

 In equilibrium both elliptic Fermi
contours are centered at $\bk=0$, and the DM coupling vanishes. A current displaces the two
contours by the same $\dk$ (Fig.~\ref{fig:mechanism}(a)). The resulting change of occupation at
the FS is odd in $\bk$, and the electrons that propagate between the two impurities
(Fig.~\ref{fig:mechanism}(b)) therefore carry an imbalance that is odd in $\bR$. Because the two
spin bands differ in their Fermi speeds and wave numbers along the bond, this imbalance differs
between the $\up$ and $\dn$ channels, and an antisymmetric coupling $\delta D_z$ appears.

\begin{figure}
\centering
\subfloat[]{\includegraphics[height=0.245\textwidth]{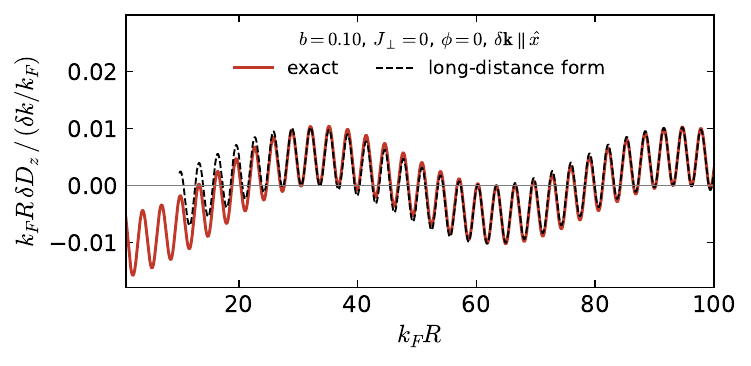}\label{fig:2a}}\hfill
\subfloat[]{\includegraphics[height=0.245\textwidth]{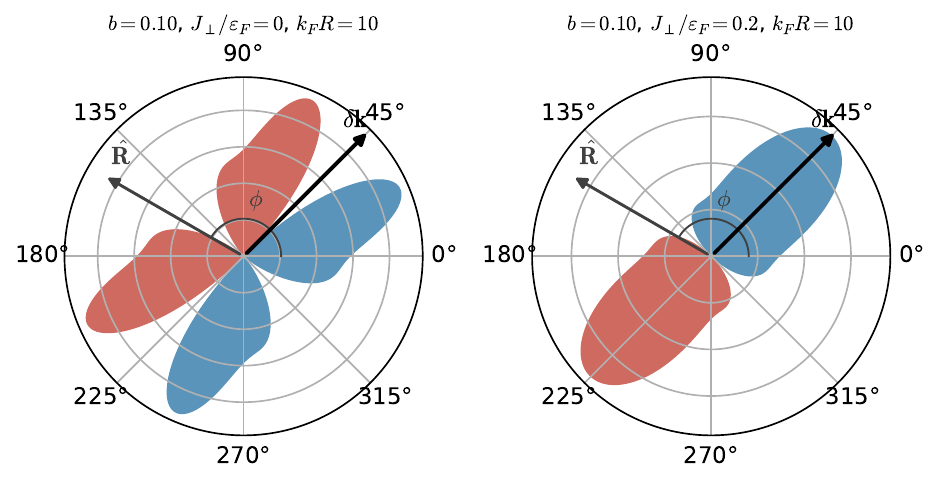}\label{fig:2b}}\\[2pt]
\subfloat[]{\includegraphics[height=0.245\textwidth]{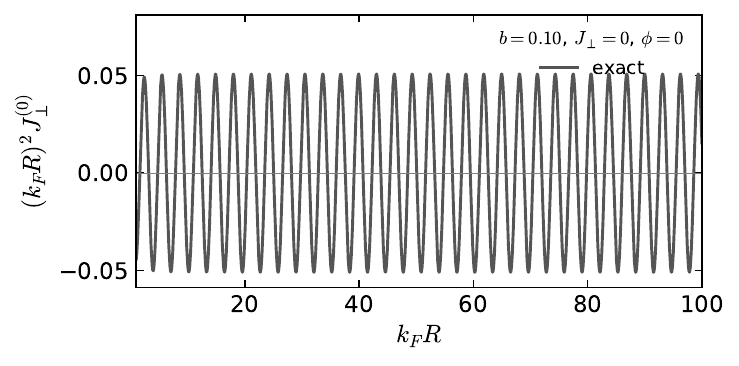}\label{fig:2c}}\hfill
\subfloat[]{\includegraphics[height=0.245\textwidth]{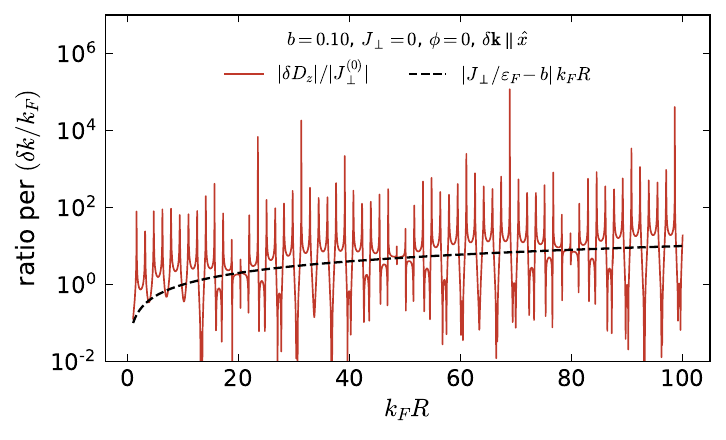}\label{fig:2d}}
\caption{Left column: (a)~exact $k_FR\,\delta D_z$ (Eq.~\eqref{eq:Dexact}) along the bond
$\phi=0$ with $\dk$ along $\hat x$ ($b=0.10$, $\Jp=0$), compared with the long-distance form,
Eq.~\eqref{eq:Dasym}; (c)~equilibrium $(k_FR)^2J_\perp^{(0)}$ for the same parameters. Right
column:
(b)~exact $k_FR\,\delta D_z$ (Eq.~\eqref{eq:Dexact}) per unit $\delta k/k_F$ as a function of the
bond angle $\phi$ at fixed $k_FR=10$, with $\dk$ along $(1,1)$, for $b=0.10$ and $\Jp=0$ (left) and
$\Jp/\eF=0.2$ (right); red, positive; blue, negative.
(d)~Ratio $|\delta D_z|/|J_\perp^{(0)}|$ per unit $\delta k/k_F$ along $\phi=0$, compared with the
envelope estimate $|\Jp/\eF-b|\,k_FR$ of Eq.~\eqref{eq:ratios}.}
\label{fig:distance}
\end{figure}

The zeros and signs of $\delta D_z$ follow directly from Eq.~\eqref{eq:Dexact}. The explicit
factor $\dk\cdot\Rh$ makes $\delta D_z$ vanish on bonds perpendicular to the drift. For $\Jp=0$,
the two spin bands have equal Fermi wave numbers and speeds along the nodal bonds
$\phi=\pm\pi/4,\pm3\pi/4$, so the bracket in Eq.~\eqref{eq:Dexact} vanishes identically there. A
rotation by $90^\circ$ interchanges the two bands and therefore reverses the sign of the bracket
at every distance. For $\dk$ along $(1,1)$, where $\dk\cdot\Rh$ is the same on bonds along $\hat x$
and $\hat y$, it follows that $\delta D_z(R\hat x)=-\delta D_z(R\hat y)$ for all $R$, which is the
$d$-wave signature of the altermagnet.

Figure~\ref{fig:distance}(b) shows the exact $\delta D_z$ as a function of the bond angle $\phi$ at
$k_FR=10$ for this geometry. In addition to the zeros identified above, $\delta D_z$ changes sign
only across the drift axis, because the bonds perpendicular to the drift are themselves nodal and
form double zeros. This pattern persists up to $k_FR\approx13$ for $b=0.10$. At larger distances,
the variation of $\Delta k(\phi)=k^F_\up(\phi)-k^F_\dn(\phi)$ with the bond direction produces
additional sign changes as a function of $\phi$.
Adding an exchange splitting changes this pattern qualitatively once $|\Jp|/\eF>|b|$. The speed
difference $v^F_\dn(\phi)-v^F_\up(\phi)$ then has the same sign for all bond directions, the
lines of equal Fermi speed disappear, and at $k_FR=10$, $\delta D_z$ changes sign only across the
line perpendicular to the drift, so that the lobes are centered on the drift axis
(Fig.~\ref{fig:distance}(b), right).

Figures~\ref{fig:distance}(a) and \ref{fig:distance}(c) contrast the current-induced and
equilibrium couplings. The envelope of $(k_FR)^2J_\perp^{(0)}$ is constant, so the equilibrium
coupling decays as $1/R^2$ and oscillates at the wave number $K(0)=k^F_\up(0)+k^F_\dn(0)\simeq2k_F$.
In contrast, $k_FR\,\delta D_z$ keeps a constant amplitude, so the DM coupling decays only as
$1/R$. Its fast oscillation at $K(0)$ is superimposed on a slow beat at the wave number
$|\Delta k(0)|=|k^F_\up(0)-k^F_\dn(0)|\approx0.10\,k_F$, with period $\approx62/k_F$, which shifts
the center of the fast oscillation. Near the extrema of the beat, $\delta D_z$ therefore keeps
essentially one sign over many fast oscillations. The long-distance form captures this structure
but converges slowly: its relative error drops below $10\%$ only for $k_FR\gtrsim65$
(Fig.~\ref{fig:checks}(b) of the Supplemental Material). Because the two couplings decay with different powers, their ratio
grows linearly with $R$ (Fig.~\ref{fig:distance}(d)). Close to the nodes of $J_\perp^{(0)}$, which
are spaced by about a quarter of the Fermi wavelength, the ratio is larger still.

\begin{figure}
\centering
\subfloat[]{\includegraphics[width=0.45\textwidth]{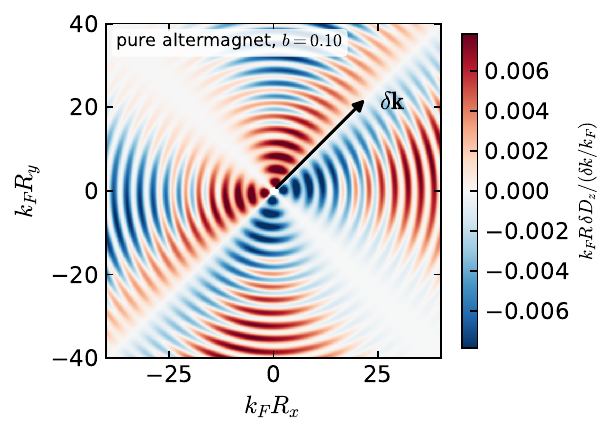}}
\hfill
\subfloat[]{\includegraphics[width=0.45\textwidth]{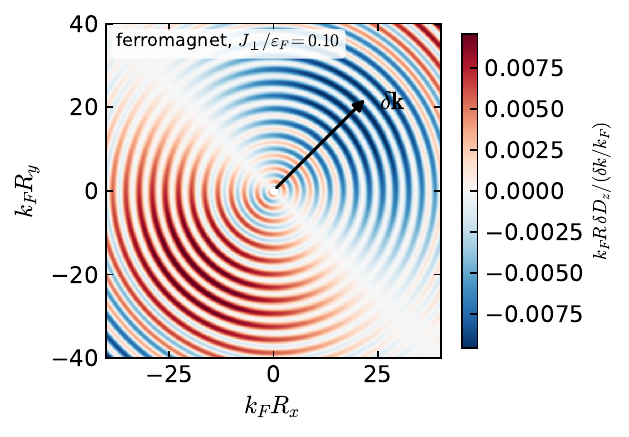}}\\[6pt]
\sbox{\figthreebox}{
\subfloat[]{\includegraphics[width=0.45\textwidth]{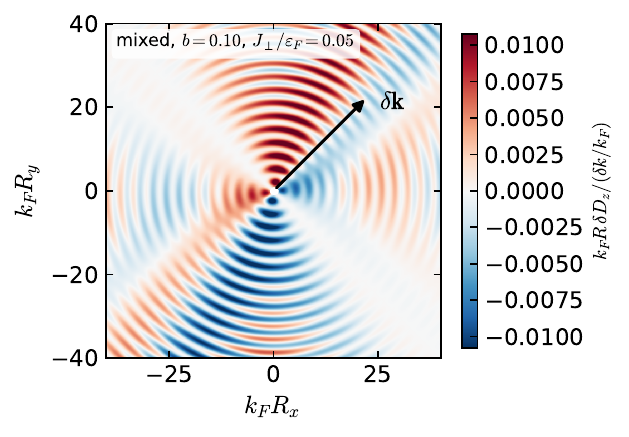}}
}%
\usebox{\figthreebox}\hfill
\subfloat[]{\includegraphics[width=0.53\textwidth,height=\ht\figthreebox,keepaspectratio]{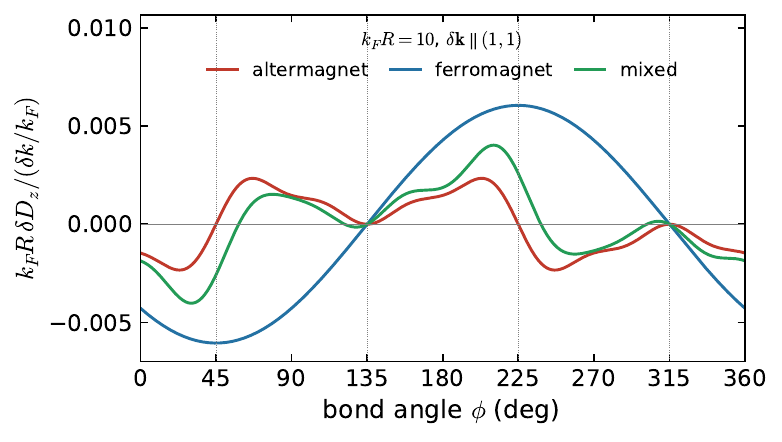}}
\caption{Real-space maps of the exact $k_FR\,\delta D_z(\bR)$ (Eq.~\eqref{eq:Dexact}) per unit
$\delta k/k_F$ for $\dk$ along $(1,1)$ (arrow). (a)~Pure altermagnet, $b=0.10$, $\Jp=0$.
(b)~Ferromagnet, $b=0$, $\Jp/\eF=0.10$. (c)~Mixed case, $b=0.10$, $\Jp/\eF=0.05$. The factor
$k_FR$ removes the $1/R$ decay. (d)~Exact $k_FR\,\delta D_z$ per unit $\delta k/k_F$ as a function
of the bond angle $\phi$ at fixed $k_FR=10$ for the three cases.}
\label{fig:maps}
\end{figure}
The angular pattern of $\delta D_z$ distinguishes an altermagnet from a ferromagnet
(Fig.~\ref{fig:maps}). In the altermagnet (Fig.~\ref{fig:maps}(a)), $\delta D_z$ vanishes on both
nodal diagonals at every distance and takes opposite values on bonds of equal length along
$\hat x$ and $\hat y$, as required by the symmetry argument above. For $k_FR\lesssim13$ the sign
changes only across the drift axis; at larger distances the direction dependence of the beat
wave number $\Delta k(\phi)$ adds further nodal lines. For $\Jp=0$, reversing the N\'eel vector
($b\to-b$) interchanges the two bands and reverses the sign of $\delta D_z$ everywhere. In the
ferromagnet (Fig.~\ref{fig:maps}(b)) the bands are isotropic, so the bracket in
Eq.~\eqref{eq:Dexact} depends only on $R$ and the angular dependence is that of $\dk\cdot\Rh$
alone: at fixed $R$, bonds ahead of and behind the drift carry opposite signs. In the mixed case
(Fig.~\ref{fig:maps}(c)) the two contributions compete, and the pattern interpolates between the
two limits. On the lines $\cos2\phi=\Jp/(b\eF)$, i.e., $\phi=\pm30^\circ$ and $\pm150^\circ$ for
the parameters used, the two bands have equal Fermi speeds, $v^F_\up(\phi)=v^F_\dn(\phi)$, but
different wave numbers; there $\delta D_z$ does not vanish identically, but it decays faster than
$1/R$.
Figure~\ref{fig:maps}(d) shows the exact $\delta D_z$ as a function of the bond angle $\phi$ at
fixed $k_FR=10$. Writing $\dk=\delta k\,(\cos\phi_d,\sin\phi_d)$, so that
$\dk\cdot\Rh=\delta k\cos(\phi-\phi_d)$, the drift points along $\phi_d=45^\circ$. In the
ferromagnet $\delta D_z$ follows $\cos(\phi-\phi_d)$ exactly, with zeros only on the bonds
perpendicular to the drift. In the altermagnet it changes sign at $45^\circ$ and $225^\circ$, the
nodal bonds along the drift, and has double zeros at $135^\circ$ and $315^\circ$, where the nodal
bonds are also perpendicular to the drift. In the mixed case additional zeros appear at
$\phi\approx58^\circ$, $121^\circ$, $238^\circ$, and $302^\circ$. They do not coincide with the lines
of equal Fermi speed, $\phi=\pm30^\circ,\pm150^\circ$, because the two bands still have different
wave numbers there.
\begin{figure}
\centering
\subfloat[]{\includegraphics[width=0.49\textwidth]{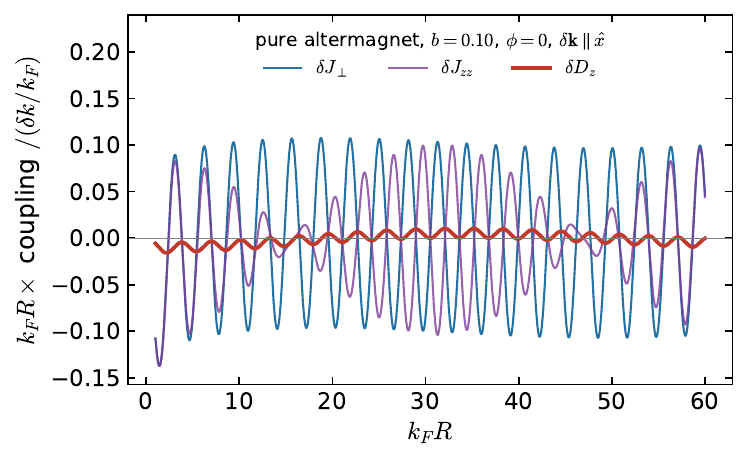}}\hfill
\subfloat[]{\includegraphics[width=0.49\textwidth]{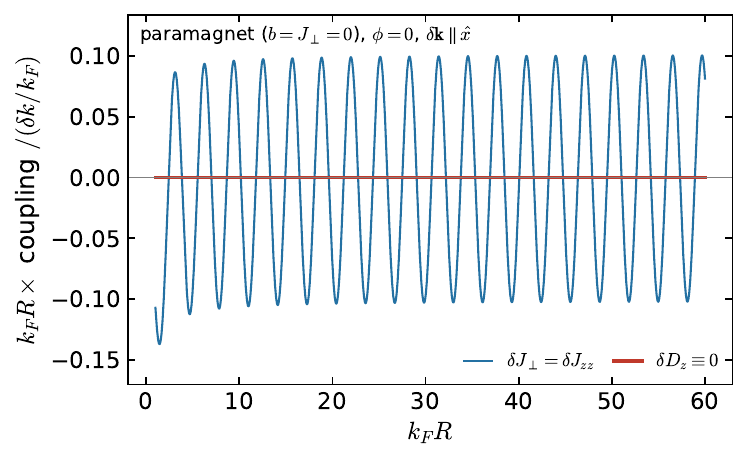}}
\caption{Exact current-induced couplings along $\phi=0$ with $\dk$ along $\hat x$, multiplied by
$k_FR$ and given per unit $\delta k/k_F$. (a)~Pure altermagnet ($b=0.10$): $\delta D_z$
(Eq.~\eqref{eq:Dexact}), $\delta J_\perp$ (Eq.~\eqref{eq:Jperpnr}), and $\delta J_{zz}$
(Eq.~\eqref{eq:Jzznr}). (b)~Paramagnet ($b=\Jp=0$): $\delta D_z\equiv0$, whereas
$\delta J_\perp=\delta J_{zz}$ remains finite.}
\label{fig:nonrecip}
\end{figure}

Figure~\ref{fig:nonrecip} compares the three current-induced couplings. The symmetric couplings
$\delta J_\perp$ and $\delta J_{zz}$ are odd in $\bR$ and therefore violate the condition
$J_{ab}(\bR)=J_{ba}(-\bR)$ satisfied by any coupling derived from a spin Hamiltonian. They exert
unequal exchange fields on the two spins and cannot be obtained from an energy function. They do
not require a spin splitting and are already present in a paramagnet (Fig.~\ref{fig:nonrecip}(b)),
where spin-rotation symmetry makes them equal, $\delta J_\perp=\delta J_{zz}$. In the altermagnet
(Fig.~\ref{fig:nonrecip}(a)) they exceed $\delta D_z$ by about an order of magnitude (rms ratio
$\approx8$--$12$ for $k_FR\le60$), but they oscillate in sign with a period of half the Fermi
wavelength. The antisymmetric coupling $\delta D_z$ satisfies $J_{ab}(\bR)=J_{ba}(-\bR)$ and is a
genuine Hamiltonian DM interaction. It requires the two spin bands to differ along the bond and
vanishes identically in a paramagnet. It is therefore the contribution specific to spin-split,
and in particular altermagnetic, conductors. Numerical checks of the exact expressions used in
this section are given in Fig.~\ref{fig:checks} of the Supplemental Material.

\section{Discussion}
\label{sec:discussion}

The current-induced DM interaction differs in origin from the spin--orbit mechanisms discussed in
Sec.~\ref{sec:intro}. Inversion symmetry is broken by the drift itself, through the factor
$\dk\cdot\Rh$, and the spin splitting $\Jp+\beta(k_x^2-k_y^2)$ distinguishes the two spin channels
without any spin--orbit coupling. Because $\sigma_z$ is conserved, the only allowed antisymmetric
component is $D_z$, along the N\'eel vector or the magnetization; the in-plane components vanish
identically (Table~\ref{tab:cab}). In equilibrium the same model gives an anisotropic but purely
symmetric RKKY tensor, $D_z^{(0)}=0$ (Eq.~\eqref{eq:J0comp}), in agreement with previous
equilibrium studies, in which a DM term appears only once Rashba coupling is
added~\cite{PhysRevB.110.054427,Yarmohammadi_2025,Kundu2025}. 
The current-induced term $\delta D_z$ is odd
in the current and can therefore be separated from any equilibrium DM coupling by reversing the
current.

Our result is consistent with earlier studies of exchange interactions out of equilibrium, which
found that an antisymmetric exchange requires spin-polarized electrons~\cite{Fransson2014} and that
a spin current flowing past two localized spins produces a DM-like interaction, whereas a charge
current alone only renormalizes the existing coupling~\cite{Onoda2006}. In a spin-split metal the
drift itself carries a spin current: the two spin bands are reoccupied differently along the bond
because their Fermi speeds and wave numbers differ. A charge current is then sufficient to
generate a DM coupling, and our results give this coupling exactly. Unlike the current-induced
modification of the DM interaction in spin--orbit-coupled multilayers~\cite{Freimuth2020,Karnad2018},
the effect requires no spin--orbit coupling and vanishes identically in a paramagnet
(Fig.~\ref{fig:nonrecip}(b)). In an altermagnet the spin polarization of the drift depends on the
direction, as in the spin-splitter effect~\cite{GonzalezHernandez2021,bose2022tilted}, and this is
the origin of the $d$-wave angular pattern of $\delta D_z$ (Figs.~\ref{fig:distance}(b) and
\ref{fig:maps}).

The symmetric corrections $\delta J_\perp$ and $\delta J_{zz}$ are of a different nature. Because
they are odd in $\bR$, the exchange field exerted by $\bS_1$ on $\bS_2$ differs from that exerted
by $\bS_2$ on $\bS_1$, and the two spins interact nonreciprocally, as in the nonreciprocal RKKY
interactions proposed for voltage-biased metals~\cite{Cherukupalli2026} and for magnetic metals
under light irradiation~\cite{Hanai2025}. Such couplings can drive time-dependent phases without
an equilibrium counterpart~\cite{Fruchart2021}, and steady, periodic, and chaotic two-spin dynamics
has already been found for localized spins coupled to charge and spin currents~\cite{Onoda2006}. In
our model these couplings are present even without spin splitting and exceed $\delta D_z$ by about
an order of magnitude (Fig.~\ref{fig:nonrecip}), but they oscillate in sign with a period of half
the Fermi wavelength. In the long-distance forms, $\delta J_{zz}$ contains only fast oscillations
(Eq.~\eqref{eq:Jzzasym}) and the slow part of $\delta J_\perp$ is proportional to
$\sin(\Delta k\,R)$ (Eq.~\eqref{eq:Jperpasym}), whereas $\delta D_z$ contains
$\cos(\Delta k\,R)$ (Eq.~\eqref{eq:Dasym}). For $R\lesssim1/|\Delta k|$, averaging over impurity
positions on the scale of the Fermi wavelength therefore suppresses the symmetric couplings much
more strongly than the DM coupling, whose average has a sign set by the bond orientation and the
direction of the current.

Our results rest on several assumptions. (i)~Propagation between the impurities is ballistic,
$R\ll\ell$. Disorder damps the real-space GFs on the scale of the mean free path
$\ell$ and cuts off the slow $1/R$ decay of the current-induced couplings beyond $\ell$.
(ii)~The response is linear in the drift, which requires $\delta k\,R\ll1$; for the low-density
two-dimensional layer estimated after Eq.~\eqref{eq:ratios}, $\delta k\,R\approx0.2$ is already at
the edge of this regime. (iii)~A single, spin-independent relaxation time gives the same drift
$\dk$ for both bands. Spin-dependent scattering makes the drift spin dependent, $\dk_\sigma$, and
weights the two terms in the bracket of Eq.~\eqref{eq:Dexact} differently; this adds a
contribution that survives even on bonds where the two bands coincide. (iv)~The results are for
$T=0$. At finite temperature the energy integral in Eq.~\eqref{eq:dJIm} averages the couplings over
a window of width $\sim k_BT$ around $\eF$. (v)~The model conserves $\sigma_z$. Rashba spin--orbit
coupling, present in gated or substrate-supported layers, breaks this conservation and adds an
equilibrium DM coupling~\cite{PhysRevB.110.054427,Kundu2025}; the coupling computed here is then
the part that is odd in the current. (vi)~The couplings are computed to second order in $\Jc$,
which neglects Kondo screening of the impurity spins.

For a bond along the current at $k_FR=20$ and a splitting factor $|\Jp/\eF-b|=0.2$, the estimates
following Eq.~\eqref{eq:ratios} give $|\delta D_z|/|J^{(0)}|\approx3\times10^{-4}$ for a metallic
film at $j=10^{11}\,{\rm A/m^2}$ and $\approx0.04$ for a low-density two-dimensional layer at a
sheet current density of $5$\,A/m. The effect is therefore most accessible in low-density,
gate-tunable spin-split conductors and in layered metallic altermagnets with $d$-wave spin
splitting, such as KV$_2$Se$_2$O~\cite{Jiang2025} and Rb$_{1-\delta}$V$_2$Te$_2$O~\cite{Zhang2025}.
Thin films of CrSb~\cite{Reimers2024} offer another platform, in which strain can convert the
$g$-wave splitting into a $d$-wave one~\cite{Karetta2025}. Pairwise exchange interactions,
including their chiral part, have been mapped for individual adatoms with spin-polarized scanning
tunneling microscopy~\cite{Zhou2010,Khajetoorians2016}. Combined with an in-plane current, such
measurements would identify $\delta D_z$ as the part of the antisymmetric exchange that reverses
with the current. Its $d$-wave fingerprint, namely opposite values on bonds along $\hat x$ and
$\hat y$ for a diagonal drift, zeros on the nodal bonds, and a sign change with the N\'eel vector,
distinguishes it from both spin--orbit-induced DM couplings and the ferromagnetic case. Joule
heating limits the usable current densities and has to be taken into account in such experiments.

Finally, in a dense ensemble of local moments the pair couplings add up to a micromagnetic DM
vector, $D_i\propto\sum_{\bR}R_i\,\delta D_z(\bR)$. For the pure $d$-wave altermagnet, the
combination of a $90^\circ$ rotation with spin reversal, together with the mirror symmetry
$x\to-x$, requires $\mathbf D\propto(\delta k_x,-\delta k_y)$, which is transverse to the drift
for a diagonal current, whereas an isotropic ferromagnet gives $\mathbf D\propto\dk$. This
suggests a current-controlled chirality of magnetic textures~\cite{Fert2017}. Its magnitude,
however, depends on how the oscillating $1/R$ tail is cut off, for example by the mean free path,
and requires a treatment beyond the ballistic limit. Together with the dynamics generated by the
nonreciprocal symmetric couplings, this is left for future work.

\section{Summary}
\label{sec:summary}

We have computed the exchange coupling between two magnetic impurities in a current-carrying,
spin-split two-dimensional metal, a $d$-wave altermagnet with an additional exchange splitting,
using Keldysh GFs. To linear order in the current, the current-induced part of the
coupling is a pure Fermi-surface term, and all its components follow in closed Bessel-function
form, exactly in the altermagnetic splitting $b$ and the exchange splitting $\Jp$
(Eqs.~\eqref{eq:Jzznr}--\eqref{eq:Dexact}).

The current generates a DM interaction $\delta D_z(\bS_1\times\bS_2)_z$ without any spin--orbit
coupling. It requires the two spin bands to differ along the bond and vanishes identically in a
paramagnet. It decays as $1/R$, one power more slowly than the equilibrium RKKY coupling, so that
its ratio to the equilibrium coupling grows linearly with $\delta k\,R$, and its fast oscillation
is superimposed on a slow beat at the wave number $\Delta k=k^F_\up-k^F_\dn$. In the pure
altermagnet $\delta D_z$ vanishes exactly on the nodal bonds and on the bonds perpendicular to the
drift, takes opposite values on bonds along $\hat x$ and $\hat y$ for a diagonal drift, and
reverses with the N\'eel vector and with the current. This $d$-wave fingerprint distinguishes an
altermagnet from a ferromagnet, in which the DM coupling changes sign only between bonds ahead of
and behind the drift.

The current also generates symmetric corrections $\delta J_\perp$ and $\delta J_{zz}$ that are odd
in $\bR$. They cannot be derived from a spin Hamiltonian, exert unequal exchange fields on the two
spins, and are present even without spin splitting. They are about an order of magnitude larger
than $\delta D_z$ but oscillate in sign on the scale of the Fermi wavelength.

The ratio of the current-induced DM coupling to the equilibrium exchange reaches a few per cent in
low-density two-dimensional conductors and about $10^{-4}$ in metallic films. Layered metallic
altermagnets combined with spin-polarized scanning tunneling microscopy offer a direct route to
observing the effect. Natural extensions of this work include disorder beyond the ballistic limit,
the micromagnetic DM interaction of dense spin ensembles, and the nonreciprocal spin dynamics
generated by the symmetric couplings.

\bibliographystyle{ieeetr}
\bibliography{references}

\begin{thebibliography}{10}

\bibitem{Ruderman1954}
M.~A. Ruderman and C.~Kittel, ``Indirect exchange coupling of nuclear magnetic
  moments by conduction electrons,'' {\em Phys. Rev.}, vol.~96, p.~99, 1954.

\bibitem{Kasuya1956}
T.~Kasuya, ``A theory of metallic ferro- and antiferromagnetism on zener's
  model,'' {\em Prog. Theor. Phys.}, vol.~16, p.~45, 1956.

\bibitem{Yosida1957}
K.~Yosida, ``Magnetic properties of cu-mn alloys,'' {\em Phys. Rev.}, vol.~106,
  p.~893, 1957.

\bibitem{Zhou2010}
L.~Zhou, J.~Wiebe, S.~Lounis, E.~Vedmedenko, F.~Meier, S.~Bl{\"u}gel, P.~H.
  Dederichs, and R.~Wiesendanger, ``Strength and directionality of surface
  {Ruderman--Kittel--Kasuya--Yosida} interaction mapped on the atomic scale,''
  {\em Nature Physics}, vol.~6, no.~3, pp.~187--191, 2010.

\bibitem{Khajetoorians2016}
A.~A. Khajetoorians, M.~Steinbrecher, M.~Ternes, M.~Bouhassoune, M.~dos
  Santos~Dias, S.~Lounis, J.~Wiebe, and R.~Wiesendanger, ``Tailoring the chiral
  magnetic interaction between two individual atoms,'' {\em Nature
  Communications}, vol.~7, p.~10620, 2016.

\bibitem{PhysRevB.58.3584}
V.~I. Litvinov and V.~K. Dugaev, ``Rkky interaction in one- and two-dimensional
  electron gases,'' {\em Phys. Rev. B}, vol.~58, pp.~3584--3585, Aug 1998.

\bibitem{PhysRevB.84.115119}
E.~Kogan, ``Rkky interaction in graphene,'' {\em Phys. Rev. B}, vol.~84,
  p.~115119, Sep 2011.

\bibitem{Sherafati2011}
M.~Sherafati and S.~Satpathy, ``Rkky interaction in graphene from the lattice
  green's function,'' {\em Phys. Rev. B}, vol.~83, p.~165425, Apr 2011.

\bibitem{Liu2009}
Q.~Liu, C.-X. Liu, C.~Xu, X.-L. Qi, and S.-C. Zhang, ``Magnetic impurities on
  the surface of a topological insulator,'' {\em Phys. Rev. Lett.}, vol.~102,
  p.~156603, Apr 2009.

\bibitem{PhysRevLett.106.097201}
J.-J. Zhu, D.-X. Yao, S.-C. Zhang, and K.~Chang, ``Electrically controllable
  surface magnetism on the surface of topological insulators,'' {\em Phys. Rev.
  Lett.}, vol.~106, p.~097201, Feb 2011.

\bibitem{Chang2015}
H.-R. Chang, J.~Zhou, S.-X. Wang, W.-Y. Shan, and D.~Xiao, ``Rkky interaction
  of magnetic impurities in dirac and weyl semimetals,'' {\em Phys. Rev. B},
  vol.~92, p.~241103, Dec 2015.

\bibitem{Hosseini2015}
M.~V. Hosseini and M.~Askari, ``Ruderman-kittel-kasuya-yosida interaction in
  weyl semimetals,'' {\em Phys. Rev. B}, vol.~92, p.~224435, Dec 2015.

\bibitem{Dzyaloshinsky1958}
I.~Dzyaloshinsky, ``A thermodynamic theory of ``weak'' ferromagnetism of
  antiferromagnetics,'' {\em J. Phys. Chem. Solids}, vol.~4, p.~241, 1958.

\bibitem{Moriya1960}
T.~Moriya, ``Anisotropic superexchange interaction and weak ferromagnetism,''
  {\em Physical Review}, vol.~120, no.~1, pp.~91--98, 1960.

\bibitem{Fert2017}
A.~Fert, N.~Reyren, and V.~Cros, ``Magnetic skyrmions: advances in physics and
  potential applications,'' {\em Nature Reviews Materials}, vol.~2, p.~17031,
  2017.

\bibitem{PhysRevLett.44.1538}
A.~Fert and P.~M. Levy, ``Role of anisotropic exchange interactions in
  determining the properties of spin-glasses,'' {\em Phys. Rev. Lett.},
  vol.~44, pp.~1538--1541, Jun 1980.

\bibitem{PhysRevB.69.121303}
H.~Imamura, P.~Bruno, and Y.~Utsumi, ``Twisted exchange interaction between
  localized spins embedded in a one- or two-dimensional electron gas with
  rashba spin-orbit coupling,'' {\em Phys. Rev. B}, vol.~69, p.~121303, Mar
  2004.

\bibitem{Bouaziz_2017}
J.~Bouaziz, M.~d.~S. Dias, A.~Ziane, M.~Benakki, S.~Blügel, and S.~Lounis,
  ``Chiral magnetism of magnetic adatoms generated by rashba electrons,'' {\em
  New Journal of Physics}, vol.~19, p.~023010, feb 2017.

\bibitem{Srivastava2018}
T.~Srivastava, M.~Schott, R.~Juge, V.~K{\v{r}}i{\v{z}}{\'a}kov{\'a},
  M.~Belmeguenai, Y.~Roussign{\'e}, A.~Bernand-Mantel, L.~Ranno, S.~Pizzini,
  S.-M. Ch{\'e}rif, A.~Stashkevich, S.~Auffret, O.~Boulle, G.~Gaudin,
  M.~Chshiev, C.~Baraduc, and H.~B{\'e}a, ``Large-voltage tuning of
  dzyaloshinskii–moriya interactions: A route toward dynamic control of
  skyrmion chirality,'' {\em Nano Letters}, vol.~18, pp.~4871--4877, 08 2018.

\bibitem{Koyama2018}
T.~Koyama, Y.~Nakatani, J.~Ieda, and D.~Chiba, ``Electric field control of
  magnetic domain wall motion via modulation of the {Dzyaloshinskii-Moriya}
  interaction,'' {\em Science Advances}, vol.~4, no.~12, p.~eaav0265, 2018.

\bibitem{PhysRevB.108.134428}
Y.~G. Semenov, X.~Xu, J.~A. Boulton, and K.~W. Kim, ``Electrical control of
  dzyaloshinskii-moriya interactions in magnetic weyl semimetals,'' {\em Phys.
  Rev. B}, vol.~108, p.~134428, Oct 2023.

\bibitem{Kikuchi2016}
T.~Kikuchi, T.~Koretsune, R.~Arita, and G.~Tatara, ``Dzyaloshinskii-moriya
  interaction as a consequence of a doppler shift due to spin-orbit-induced
  intrinsic spin current,'' {\em Phys. Rev. Lett.}, vol.~116, p.~247201, Jun
  2016.

\bibitem{Freimuth2020}
F.~Freimuth, S.~Bl{\"u}gel, and Y.~Mokrousov, ``Dynamical and current-induced
  dzyaloshinskii-moriya interaction: Role for damping, gyromagnetism, and
  current-induced torques in noncollinear magnets,'' {\em Phys. Rev. B},
  vol.~102, p.~245411, Dec 2020.

\bibitem{Karnad2018}
G.~V. Karnad, F.~Freimuth, E.~Martinez, R.~Lo~Conte, G.~Gubbiotti, T.~Schulz,
  S.~Senz, B.~Ocker, Y.~Mokrousov, and M.~Kl{\"a}ui, ``Modification of
  dzyaloshinskii-moriya-interaction-stabilized domain wall chirality by driving
  currents,'' {\em Phys. Rev. Lett.}, vol.~121, p.~147203, Oct 2018.

\bibitem{smejkal2022beyond}
L.~{\ifmmode \check{S}\else \v{S}\fi{}}mejkal, J.~Sinova, and T.~Jungwirth,
  ``Beyond conventional ferromagnetism and antiferromagnetism: A phase with
  nonrelativistic spin and crystal rotation symmetry,'' {\em Phys. Rev. X},
  vol.~12, p.~031042, Sep 2022.

\bibitem{smejkal2022emerging}
L.~{\ifmmode \check{S}\else \v{S}\fi{}}mejkal, J.~Sinova, and T.~Jungwirth,
  ``Emerging research landscape of altermagnetism,'' {\em Phys. Rev. X},
  vol.~12, p.~040501, Dec 2022.

\bibitem{Krempasky2024_Nature}
J.~Krempask{\'y}, L.~{\v{S}}mejkal, S.~W. D'Souza, M.~Hajlaoui, G.~Springholz,
  K.~Uhl{\'i}{\v{r}}ov{\'a}, F.~Alarab, P.~C. Constantinou, V.~Strocov,
  D.~Usanov, W.~R. Pudelko, R.~Gonz{\'a}lez-Hern{\'a}ndez, A.~Birk~Hellenes,
  Z.~Jansa, H.~Reichlov{\'a}, Z.~{\v{S}}ob{\'a}{\v{n}}, R.~D.
  Gonzalez~Betancourt, P.~Wadley, J.~Sinova, D.~Kriegner, J.~Min{\'a}r, J.~H.
  Dil, and T.~Jungwirth, ``Altermagnetic lifting of kramers spin degeneracy,''
  {\em Nature}, vol.~626, pp.~517--522, Feb 2024.

\bibitem{Amin2024}
O.~J. Amin, A.~Dal~Din, E.~Golias, Y.~Niu, A.~Zakharov, S.~C. Fromage, C.~J.~B.
  Fields, S.~L. Heywood, R.~B. Cousins, F.~Maccherozzi, J.~Krempask{\'y}, J.~H.
  Dil, D.~Kriegner, B.~Kiraly, R.~P. Campion, A.~W. Rushforth, K.~W. Edmonds,
  S.~S. Dhesi, L.~{\v{S}}mejkal, T.~Jungwirth, and P.~Wadley, ``Nanoscale
  imaging and control of altermagnetism in mnte,'' {\em Nature}, vol.~636,
  pp.~348--353, Dec 2024.

\bibitem{Reimers2024}
S.~Reimers, L.~Odenbreit, L.~{\v{S}}mejkal, V.~N. Strocov, P.~Constantinou,
  A.~B. Hellenes, R.~Jaeschke~Ubiergo, W.~H. Campos, V.~K. Bharadwaj,
  A.~Chakraborty, T.~Denneulin, W.~Shi, R.~E. Dunin-Borkowski, S.~Das,
  M.~Kl{\"a}ui, J.~Sinova, and M.~Jourdan, ``Direct observation of
  altermagnetic band splitting in crsb thin films,'' {\em Nature
  Communications}, vol.~15, p.~2116, Mar 2024.

\bibitem{Jiang2025}
B.~Jiang, M.~Hu, J.~Bai, Z.~Song, C.~Mu, G.~Qu, W.~Li, W.~Zhu, H.~Pi, Z.~Wei,
  Y.-J. Sun, Y.~Huang, X.~Zheng, Y.~Peng, L.~He, S.~Li, J.~Luo, Z.~Li, G.~Chen,
  H.~Li, H.~Weng, and T.~Qian, ``A metallic room-temperature d-wave
  altermagnet,'' {\em Nature Physics}, vol.~21, pp.~754--759, May 2025.

\bibitem{Zhang2025}
F.~Zhang, X.~Cheng, Z.~Yin, C.~Liu, L.~Deng, Y.~Qiao, Z.~Shi, S.~Zhang, J.~Lin,
  Z.~Liu, M.~Ye, Y.~Huang, X.~Meng, C.~Zhang, T.~Okuda, K.~Shimada, S.~Cui,
  Y.~Zhao, G.-H. Cao, S.~Qiao, J.~Liu, and C.~Chen, ``Crystal-symmetry-paired
  spin--valley locking in a layered room-temperature metallic altermagnet
  candidate,'' {\em Nature Physics}, vol.~21, pp.~760--767, May 2025.

\bibitem{GonzalezHernandez2021}
R.~Gonz{\'a}lez-Hern{\'a}ndez, L.~{\ifmmode \check{S}\else \v{S}\fi{}}mejkal,
  K.~V{\'y}born{\'y}, Y.~Yahagi, J.~Sinova, T.~Jungwirth, and J.~{\ifmmode
  \check{Z}\else \v{Z}\fi{}}elezn{\'y}, ``Efficient electrical spin splitter
  based on nonrelativistic collinear antiferromagnetism,'' {\em Phys. Rev.
  Lett.}, vol.~126, p.~127701, Mar 2021.

\bibitem{bose2022tilted}
A.~Bose, N.~J. Schreiber, R.~Jain, D.-F. Shao, H.~P. Nair, J.~Sun, X.~S. Zhang,
  D.~A. Muller, E.~Y. Tsymbal, D.~G. Schlom, and D.~C. Ralph, ``Tilted spin
  current generated by the collinear antiferromagnet ruthenium dioxide,'' {\em
  Nature Electronics}, vol.~5, pp.~267--274, May 2022.

\bibitem{PhysRevB.110.054427}
M.~Amundsen, A.~Brataas, and J.~Linder, ``Rkky interaction in rashba
  altermagnets,'' {\em Phys. Rev. B}, vol.~110, p.~054427, Aug 2024.

\bibitem{Lee2025}
Y.-L. Lee, ``Magnetic impurities in an altermagnetic metal,'' {\em The European
  Physical Journal B}, vol.~98, p.~43, Mar 2025.

\bibitem{Yarmohammadi_2025}
M.~Yarmohammadi, U.~Zülicke, J.~Berakdar, J.~Linder, and J.~K. Freericks,
  ``Anisotropic light-tailored {RKKY} interaction in two-dimensional $d$-wave
  altermagnets,'' {\em Physical Review B}, vol.~111, June 2025.

\bibitem{Kundu2025}
A.~Kundu, ``{RKKY} interaction mediated by a spin-polarized {2D} electron gas
  with {Rashba} and altermagnetic coupling,'' 2025.

\bibitem{Schwabe1996}
N.~F. Schwabe, R.~J. Elliott, and N.~S. Wingreen,
  ``Ruderman-kittel-kasuya-yosida interaction across a tunneling junction out
  of equilibrium,'' {\em Phys. Rev. B}, vol.~54, pp.~12953--12968, Nov 1996.

\bibitem{Fransson2010}
J.~Fransson, ``Dynamical exchange interaction between localized spins out of
  equilibrium,'' {\em Phys. Rev. B}, vol.~82, p.~180411(R), Nov 2010.

\bibitem{Fransson2014}
J.~Fransson, J.~Ren, and J.-X. Zhu, ``Electrical and thermal control of
  magnetic exchange interactions,'' {\em Phys. Rev. Lett.}, vol.~113,
  p.~257201, Dec 2014.

\bibitem{Onoda2006}
M.~Onoda and N.~Nagaosa, ``Dynamics of localized spins coupled to the
  conduction electrons with charge and spin currents,'' {\em Phys. Rev. Lett.},
  vol.~96, p.~066603, Feb 2006.

\bibitem{Cherukupalli2026}
A.~Cherukupalli, G.~Refael, and J.~F. Steiner, ``Nonreciprocal {RKKY}
  interactions and chiral phases in voltage-biased metals.'' APS March Meeting
  2026, abstract, 2026.

\bibitem{Hanai2025}
R.~Hanai, D.~Ootsuki, and R.~Tazai, ``Photoinduced non-reciprocal magnetism,''
  {\em Nature Communications}, vol.~16, p.~8195, Sep 2025.

\bibitem{Keldysh1965}
L.~V. Keldysh, ``Diagram technique for nonequilibrium processes,'' {\em Sov.
  Phys. JETP}, vol.~20, p.~1018, 1965.

\bibitem{Rammer2007}
J.~Rammer, {\em Quantum Field Theory of Non-equilibrium States}.
\newblock Cambridge: Cambridge University Press, 2007.

\bibitem{HaugJauho2008}
H.~Haug and A.-P. Jauho, {\em Quantum Kinetics in Transport and Optics of
  Semiconductors}.
\newblock Berlin: Springer, 2nd~ed., 2008.

\bibitem{Kamenev2011}
A.~Kamenev, {\em Field Theory of Non-Equilibrium Systems}.
\newblock Cambridge: Cambridge University Press, 2011.

\bibitem{Stefanucci2013}
G.~Stefanucci and R.~van Leeuwen, {\em Nonequilibrium Many-Body Theory of
  Quantum Systems: A Modern Introduction}.
\newblock Cambridge: Cambridge University Press, 2013.

\bibitem{Fruchart2021}
M.~Fruchart, R.~Hanai, P.~B. Littlewood, and V.~Vitelli, ``Non-reciprocal phase
  transitions,'' {\em Nature}, vol.~592, p.~363, 2021.

\bibitem{Karetta2025}
B.~Karetta, X.~H. Verbeek, R.~Jaeschke-Ubiergo, L.~{\v{S}}mejkal, and
  J.~Sinova, ``Strain-controlled $g$- to $d$-wave transition in altermagnetic
  {CrSb},'' {\em Phys. Rev. B}, vol.~112, p.~094454, 2025.

\end{thebibliography}

\clearpage
\onecolumngrid            
\section*{Supplemental Material}

\setcounter{figure}{0}  \renewcommand{\thefigure}{S\arabic{figure}}
\setcounter{table}{0}   \renewcommand{\thetable}{S\arabic{table}}
\setcounter{equation}{0}\renewcommand{\theequation}{S\arabic{equation}}
\setcounter{section}{0} \renewcommand{\thesection}{S\arabic{section}}

\title{Supplemental Material for ``Current-driven chiral RKKY interaction in altermagnets''}
\author{A.~Kundu}
\affiliation{Asia Pacific Center for Theoretical Physics, POSTECH, Pohang 37673, Korea.}
\maketitle
\section{Occupation number: out-of-equillibrium}
 In the drifting Fermi sea the occupation of state
$\bk\sigma$ is the equilibrium occupation evaluated at the shifted momentum,
\begin{equation}
f_\sigma(\bk)=f_0\big(\varepsilon_\sigma(\bk-\dk)\big),\qquad
f_0(\varepsilon)=\frac{1}{e^{(\varepsilon-\eF)/k_BT}+1},\qquad
\dk=-\frac{e\tau\mathbf E}{\hbar}.
\label{eq:fshift}
\end{equation}
The drift $\dk$ follows from the Boltzmann equation in the relaxation-time approximation.
For electrons of charge $-e$ the force is $\hbar\dot{\bk}=-e\mathbf E$, and in the
homogeneous steady state
$-\frac{e}{\hbar}\mathbf E\cdot\nabla_{\bk}f_\sigma=-\frac{f_\sigma-f_0}{\tau}$. To first order in
$\mathbf E$ this gives
$f_\sigma(\bk)=f_0+({e\tau}/{\hbar})\mathbf E\cdot\nabla_{\bk}f_0\simeq f_0\big(\varepsilon_\sigma(\bk-\dk)\big)$
with $\dk=-e\tau\mathbf E/\hbar$: the Fermi sea is rigidly displaced by the momentum gained
from the field during one relaxation time. A spin-independent $\tau$ is assumed, so both
spin bands drift by the same $\dk$.
For $|\dk|\ll k_F$, expand the band energy to first order in $\dk$:
\begin{equation}
\varepsilon_\sigma(\bk-\dk)=\varepsilon_\sigma(\bk)-\dk\cdot\nabla_{\bk}\varepsilon_\sigma(\bk)+O(\delta k^2).
\end{equation}
Then expand $f_0$ about $\varepsilon_\sigma(\bk)$:
\begin{equation}
\begin{array}{rcl}
f_\sigma(\bk)&=&f_0\big(\varepsilon_\sigma(\bk)-\dk\cdot\nabla_{\bk}\varepsilon_\sigma\big)\\
&=&f_0\big(\varepsilon_\sigma(\bk)\big)-\big(\dk\cdot\nabla_{\bk}\varepsilon_\sigma\big)\,f_0'\big(\varepsilon_\sigma(\bk)\big)+O(\delta k^2).
\end{array}
\end{equation}
Hence
\begin{equation}
\begin{array}{rcl}
f_\sigma(\bk)&=&f_0\big(\varepsilon_\sigma(\bk)\big)+\delta f_\sigma(\bk),\\
\delta f_\sigma(\bk)&=&\big[-f_0'\big(\varepsilon_\sigma(\bk)\big)\big]\,\big(\dk\cdot\nabla_{\bk}\varepsilon_\sigma(\bk)\big)
=\big[-f_0'\big(\varepsilon_\sigma(\bk)\big)\big]\,\hbar\,\dk\cdot\mathbf v_\sigma(\bk),
\end{array}
\label{eq:dfk}
\end{equation}
with the group velocity $\mathbf v_\sigma(\bk)=\hbar^{-1}\nabla_{\bk}\varepsilon_\sigma(\bk)$. For the
bands of Eq.~\mref{eq:H},
\begin{equation}
\dk\cdot\nabla_{\bk}\varepsilon_\sigma(\bk)=\hbar^2\Big(\frac{\delta k_x\,k_x}{m_{x\sigma}}+\frac{\delta k_y\,k_y}{m_{y\sigma}}\Big).
\end{equation}
Inserting $f_\sigma(\bk)$ into $G^<_\sigma(\bk,\varepsilon)=i f_\sigma(\bk)A_\sigma(\bk,\varepsilon)$ and
using $\varepsilon=\varepsilon_\sigma(\bk)$ on shell, $f_0'(\varepsilon_\sigma(\bk))\to f_0'(\varepsilon)$,
gives $G^<_{\sigma,{\rm eq}}=if_0(\varepsilon)A_\sigma$ and
\begin{equation}
\delta G^<_\sigma(\bk,\varepsilon)=i\,\big[-f_0'(\varepsilon)\big]\,\big(\dk\cdot\nabla_{\bk}\varepsilon_\sigma(\bk)\big)\,A_\sigma(\bk,\varepsilon),
\end{equation}
which is Eq.~\eqref{eq:dGlk}.

\section{GF in k-space}
Write
$G^<_\sigma(\bk,\varepsilon)=G^<_{\sigma,{\rm eq}}(\bk,\varepsilon)+\delta G^<_\sigma(\bk,\varepsilon)$,
with the equilibrium part
$G^<_{\sigma,{\rm eq}}=-f_0(\varepsilon)[G^R_\sigma-G^A_\sigma]=i f_0(\varepsilon)A_\sigma(\bk,\varepsilon)$
and the current-induced correction
$\delta G^<_\sigma=i[-f_0'(\varepsilon)][\dk\cdot\nabla_{\bk}\varepsilon_\sigma(\bk)]A_\sigma(\bk,\varepsilon)$.
Here $A_\sigma(\bk,\varepsilon)=2\pi\delta(\varepsilon-\varepsilon_\sigma(\bk))$ is the spectral
function, $f_0$ is the Fermi function with chemical potential $\eF$, and
$\dk=-e\tau\mathbf E/\hbar$ is the drift of the Fermi sea. Since $-f_0'$ is peaked at $\eF$,
$\delta G^<_\sigma\propto\dk$ is confined to the Fermi surface and describes its shift.
More details ($k$-space functions):\\
\begin{equation}
\begin{array}{rcl}
f_\sigma(\bk)&=&f_0\big(\varepsilon_\sigma(\bk)\big)
-\big[\dk\cdot\nabla_{\bk}\varepsilon_\sigma(\bk)\big]\,f_0'\big(\varepsilon_\sigma(\bk)\big)+O(\delta k^2),\\
G^<_\sigma(\bk,\varepsilon)&=&i\,f_\sigma(\bk)\,A_\sigma(\bk,\varepsilon)
=G^<_{\sigma,{\rm eq}}(\bk,\varepsilon)+\delta G^<_\sigma(\bk,\varepsilon),\\
G^<_{\sigma,{\rm eq}}(\bk,\varepsilon)&=&i\,f_0(\varepsilon)\,A_\sigma(\bk,\varepsilon)\\
&=&-f_0(\varepsilon)\big[G^R_\sigma(\bk,\varepsilon)-G^A_\sigma(\bk,\varepsilon)\big],\\
\delta G^<_\sigma(\bk,\varepsilon)&=&i\,\big[-f_0'(\varepsilon)\big]\,
\big[\dk\cdot\nabla_{\bk}\varepsilon_\sigma(\bk)\big]\,A_\sigma(\bk,\varepsilon).
\end{array}
\label{eq:Glsplit}
\end{equation}
with
\begin{equation}
\begin{array}{rcl}
G^{R}_\sigma(\bk,\varepsilon)&=&\dfrac{1}{\varepsilon-\varepsilon_\sigma(\bk)+i0},\\
G^{A}_\sigma(\bk,\varepsilon)&=&\dfrac{1}{\varepsilon-\varepsilon_\sigma(\bk)-i0},\\
A_\sigma(\bk,\varepsilon)&=&i\big[G^R_\sigma(\bk,\varepsilon)-G^A_\sigma(\bk,\varepsilon)\big]
=2\pi\,\delta\big(\varepsilon-\varepsilon_\sigma(\bk)\big),\\
\varepsilon_\sigma(\bk)&=&\dfrac{\hbar^2k_x^2}{2m_{x\sigma}}+\dfrac{\hbar^2k_y^2}{2m_{y\sigma}}+\sigma\Jp,\\
\dk\cdot\nabla_{\bk}\varepsilon_\sigma(\bk)&=&\hbar^2\Big(\dfrac{\delta k_x\,k_x}{m_{x\sigma}}+\dfrac{\delta k_y\,k_y}{m_{y\sigma}}\Big),\\
f_0(\varepsilon)&=&\dfrac{1}{e^{(\varepsilon-\eF)/k_BT}+1},\\
-f_0'(\varepsilon)&\xrightarrow{T=0}&\delta(\varepsilon-\eF),\\
\dk&=&-\dfrac{e\tau\mathbf E}{\hbar}.
\end{array}
\label{eq:Glsplit2}
\end{equation}

\section{ Retarded Green's function in real space}
In a homogeneous steady state every function depends on $\mathbf r_1-\mathbf r_2$ and
$t_1-t_2$ only, and we use
\begin{equation}
G(\bR,t)=\int\!\frac{d^2k}{(2\pi)^2}\int\!\frac{d\varepsilon}{2\pi\hbar}\;
e^{i\bk\cdot\bR-i\varepsilon t/\hbar}\,G(\bk,\varepsilon) .
\label{eq:FTconvo}
\end{equation}
\textit{Assumptions:} (i) $m_{x\sigma},m_{y\sigma}>0$ ($|b|<1$); (ii) $\varepsilon>\sigma\Jp$; (iii) $R\neq0$.
By Eqs.~\eqref{eq:FTconvo} and \eqref{eq:Glsplit},
\begin{equation}
G^R_\sigma(\bR,\varepsilon)=\int\!\frac{d^2k}{(2\pi)^2}\,\frac{e^{i\bk\cdot\bR}}{\varepsilon-\varepsilon_\sigma(\bk)+i0}.
\end{equation}
With the rescaling  above, $k_j=\sqrt{m_{j\sigma}}\,p_{\sigma,j}$,
\begin{equation}
\begin{array}{rcl}
G^R_\sigma(\bR,\varepsilon)&=&M\displaystyle\int\!\frac{d^2p_\sigma}{(2\pi)^2}\,
\frac{e^{i\mathbf p_\sigma\cdot\bm\rho_\sigma}}{\varepsilon-\sigma\Jp-\hbar^2p_\sigma^2/2+i0}\\
&=&\dfrac{2M}{\hbar^2}\displaystyle\int\!\frac{d^2p_\sigma}{(2\pi)^2}\,
\frac{e^{i\mathbf p_\sigma\cdot\bm\rho_\sigma}}{\kappa_\sigma^2-p_\sigma^2+i0},
\qquad \kappa_\sigma=\frac{\sqrt{2(\varepsilon-\sigma\Jp)}}{\hbar}.
\end{array}
\end{equation}
 In polar coordinates about $\bm\rho_\sigma$,
$\int_0^{2\pi}d\vartheta_\sigma\,e^{ip_\sigma\rho_\sigma\cos\vartheta_\sigma}=2\pi J_0(p_\sigma\rho_\sigma)$, so
\begin{equation}
G^R_\sigma(\bR,\varepsilon)=\frac{M}{\pi\hbar^2}\,\mathcal I_\sigma,\qquad
\mathcal I_\sigma\equiv\int_0^\infty\!dp_\sigma\;\frac{p_\sigma\,J_0(p_\sigma\rho_\sigma)}{\kappa_\sigma^2-p_\sigma^2+i0}.
\end{equation}
 Write $J_0=\tfrac12\big[H_0^{(1)}+H_0^{(2)}\big]$. With
$H_0^{(2)}(x)=-H_0^{(1)}(e^{i\pi}x)$ for $x>0$ and the substitution $q_\sigma=-p_\sigma$ in the $H_0^{(2)}$
part ($p_\sigma\,dp_\sigma=q_\sigma\,dq_\sigma$),
\begin{equation}
\int_0^\infty\!dp_\sigma\,\frac{p_\sigma\,H_0^{(2)}(p\rho_\sigma)}{\kappa_\sigma^2-p_\sigma^2+i0}
=\int_{-\infty}^{0}\!dq_\sigma\,\frac{q_\sigma\,H_0^{(1)}(q_\sigma\rho_\sigma)}{\kappa_\sigma^2-q_\sigma^2+i0},
\end{equation}
where $q<0$ is taken just above the branch cut of $H_0^{(1)}$ on the negative real axis. Hence
\begin{equation}
\mathcal I_\sigma=\frac12\int_{-\infty}^{\infty}\!dq_\sigma\;\frac{q_\sigma\,H_0^{(1)}(q_\sigma\rho_\sigma)}{\kappa_\sigma^2-q_\sigma^2+i0}.
\end{equation}
Since $\kappa_\sigma^2+i0=(\kappa_\sigma+i0)^2$,
\begin{equation}
\frac{1}{\kappa_\sigma^2-q_\sigma^2+i0}=-\frac{1}{(q_\sigma-\kappa_\sigma-i0)(q_\sigma+\kappa_\sigma+i0)},
\end{equation}
with one pole at $q_\sigma=\kappa_\sigma+i0$ (upper half plane) and one at $q_\sigma=-\kappa_\sigma-i0$
(lower half plane).
For $|q_\sigma|\to\infty$ with
${\rm Im}\,q_\sigma>0$, $H_0^{(1)}(q_\sigma\rho_\sigma)\simeq\sqrt{2/(\pi q\rho_\sigma)}\,e^{i(q_\sigma\rho_\sigma-\pi/4)}$
decays exponentially, so the large arc gives zero (Jordan's lemma). The branch cut lies below
the contour and is not crossed. Only the pole at $q_\sigma=\kappa_\sigma$ is enclosed:
\begin{equation}
\begin{array}{rcl}
\displaystyle\int_{-\infty}^{\infty}\!dq_\sigma\;\frac{q_\sigma\,H_0^{(1)}(q_\sigma\rho_\sigma)}{\kappa_\sigma^2-q_\sigma^2+i0}
&=&2\pi i\;{\rm Res}_{q_\sigma=\kappa_\sigma}\Big[-\dfrac{q_\sigma\,H_0^{(1)}(q_\sigma\rho_\sigma)}{(q_\sigma-\kappa_\sigma)(q_\sigma+\kappa_\sigma)}\Big]\\
&=&2\pi i\,\Big[-\dfrac{\kappa_\sigma\,H_0^{(1)}(\kappa_\sigma\rho_\sigma)}{2\kappa_\sigma}\Big]
=-i\pi\,H_0^{(1)}(\kappa_\sigma\rho_\sigma),
\end{array}
\end{equation}
so that $\mathcal I_\sigma=-\tfrac{i\pi}{2}\,H_0^{(1)}(\kappa_\sigma\rho_\sigma)$.
Using  $\rho_\sigma=R\sqrt{m_\sigma(\phi)}$, $m_\sigma(\phi)=m_{x\sigma}\cos^2\phi+m_{y\sigma}\sin^2\phi$, and $k_\sigma(\phi,\varepsilon)=\sqrt{2m_\sigma(\phi)(\varepsilon-\sigma\Jp)}/\hbar$ is the wave number along the bond direction $\phi$ we get$,\kappa_\sigma\rho_\sigma=k_\sigma(\phi,\varepsilon)R$  and, 
\begin{equation}
G^R_\sigma(\bR,\varepsilon)
=-\frac{iM}{2\hbar^2}\,H_0^{(1)}\big(k_\sigma(\phi,\varepsilon)R\big).
\label{gr-epsilon}
\end{equation}
\subsection{Properties of the Green's function}
For even bands the retarded Green's function is even in $\bR$, and we write
\begin{equation}
G^R_\sigma(-\bR,\varepsilon)=G^R_\sigma(\bR,\varepsilon)\equiv G_\sigma(\varepsilon),\qquad
G_\sigma(\varepsilon)=-\frac{iM}{2\hbar^2}\,H_0^{(1)}\big(k_\sigma(\phi,\varepsilon)R\big).
\label{gsigma-epsilon}
\end{equation}

\section{Energy integration and zeroth-order-RKKY}\label{s-zeroth-order-RKKY}
With Table~\ref{tab:cab},
\begin{equation}
\begin{array}{rcl}
{\rm Tr}\big[\sigma_x\,G^R(\bR,\varepsilon)\,\sigma_x\,G^R(-\bR,\varepsilon)\big]&=&G_\dn(\varepsilon)\,G_\up(\varepsilon)+G_\up(\varepsilon)\,G_\dn(\varepsilon)=2\,G_\up(\varepsilon)\,G_\dn(\varepsilon),\\
{\rm Tr}\big[\sigma_z\,G^R(\bR,\varepsilon)\,\sigma_z\,G^R(-\bR,\varepsilon)\big]&=&G_\up^2(\varepsilon)+G_\dn^2(\varepsilon),\\
{\rm Tr}\big[\sigma_x\,G^R(\bR,\varepsilon)\,\sigma_y\,G^R(-\bR,\varepsilon)\big]&=&i\,\big[G_\dn(\varepsilon)\,G_\up(\varepsilon)-G_\up(\varepsilon)\,G_\dn(\varepsilon)\big]=0 .
\end{array}
\end{equation}
The $xy$ channel vanishes because, with $G^R_\sigma(-\bR,\varepsilon)=G^R_\sigma(\bR,\varepsilon)$,
both terms reduce to the same product $G_\up(\varepsilon)G_\dn(\varepsilon)$; hence there is no
equilibrium DM term. The mixed components $(x,z)$, $(y,z)$ vanish by Table~\ref{tab:cab}:
\begin{equation}
\begin{array}{rcl}
J_\perp^{(0)}\equiv J^{(0)}_{xx}=J^{(0)}_{yy}&=&-\dfrac{2\Jc^2}{\pi}\,{\rm Im}\displaystyle\int_{-\infty}^{\eF}\!d\varepsilon\;G_\up(\varepsilon)\,G_\dn(\varepsilon),\\
J_{zz}^{(0)}&=&-\dfrac{\Jc^2}{\pi}\,{\rm Im}\displaystyle\int_{-\infty}^{\eF}\!d\varepsilon\;\big[G_\up^2(\varepsilon)+G_\dn^2(\varepsilon)\big],\\
D_z^{(0)}&=&-J^{(0)}_{xy}=0 .
\end{array}
\label{eq:J0comp}
\end{equation}
 Using $G_\sigma G_{\sigma'}=-\frac{M M_{\sigma'}}{4\hbar^4}\,H_0^{(1)}(k_\sigma R)\,H_0^{(1)}(k_{\sigma'}R)$,
\begin{equation}
\begin{array}{rcl}
J_\perp^{(0)}(\bR)&=&\dfrac{\Jc^2M^2}{2\pi\hbar^4}\,{\rm Im}\displaystyle\int_{-\infty}^{\eF}\!d\varepsilon\;
H_0^{(1)}\big(k_\up(\phi,\varepsilon)R\big)\,H_0^{(1)}\big(k_\dn(\phi,\varepsilon)R\big),\\
J_{zz}^{(0)}(\bR)&=&\dfrac{\Jc^2 M^2}{4\pi\hbar^4}\displaystyle\sum_\sigma\,{\rm Im}\int_{-\infty}^{\eF}\!d\varepsilon\;
\Big[H_0^{(1)}\big(k_\sigma(\phi,\varepsilon)R\big)\Big]^2 .
\end{array}
\label{eq:J0H}
\end{equation}
We assume that the Fermi energy lies well above both band bottoms, $\eF>|\Jp|$, so that both
spin bands are occupied and $k^F_\sigma(\phi)$ is real for both spins.
The energy integrals in Eq.~\eqref{eq:J0H} formally start at
$-\infty$, but only energies at which states exist contribute. Below the bottom of band
$\sigma$, $\varepsilon<\sigma\Jp$, $k_\sigma$ is imaginary and
$G_\sigma=-(M/\pi\hbar^2)K_0(|\kappa_\sigma|\rho_\sigma)$ is real. A product of two real
propagators has no imaginary part. Hence the $zz$ integral starts at the band bottom
$\sigma\Jp$, and the $\perp$ integral starts at the lower band bottom $-|\Jp|$. For
$\Jp\neq0$, the strip $-|\Jp|<\varepsilon<|\Jp|$ between the two band bottoms lies below $\eF$
and contributes to $J_\perp^{(0)}$: there only one band has states, and the product is complex.

From $k_\sigma(\phi,\varepsilon)$ defined before  Eq.~\eqref{gr-epsilon},
\begin{equation}
\begin{array}{rcl}
k_\sigma^2(\phi,\varepsilon)&=&\dfrac{2m_\sigma(\phi)\,(\varepsilon-\sigma\Jp)}{\hbar^2}
\quad\Longrightarrow\quad
\varepsilon=\sigma\Jp+\dfrac{\hbar^2k_\sigma^2(\phi,\varepsilon)}{2m_\sigma(\phi)},\\
d\varepsilon&=&\dfrac{\hbar^2}{m_\sigma(\phi)}\,k_\sigma(\phi,\varepsilon)\,dk_\sigma(\phi,\varepsilon).
\end{array}
\end{equation}
The limits are
\begin{equation}
\varepsilon=\sigma\Jp\;\Longrightarrow\;k_\sigma(\phi,\sigma\Jp)=0,\qquad
\varepsilon=\eF\;\Longrightarrow\;k_\sigma(\phi,\eF)=k^F_\sigma(\phi).
\end{equation}

\paragraph{(a) $J_{zz}^{(0)}$: exact for all $\beta$, $\Jp$.} We change the integration variable
from $\varepsilon$ to $q=k_\sigma(\phi,\varepsilon)$, so that
$\varepsilon=\sigma\Jp+\hbar^2q^2/2m_\sigma(\phi)$ and $d\varepsilon=(\hbar^2/m_\sigma(\phi))\,q\,dq$,
with $q=0$ at the band bottom $\varepsilon=\sigma\Jp$ and $q=k^F_\sigma(\phi)$ at $\varepsilon=\eF$.
With ${\rm Im}\big[H_0^{(1)}(x)\big]^2={\rm Im}(J_0+iY_0)^2=2J_0(x)Y_0(x)$,
\begin{equation}
{\rm Im}\int_{\sigma\Jp}^{\eF}\!d\varepsilon\,\Big[H_0^{(1)}\big(k_\sigma(\phi,\varepsilon)R\big)\Big]^2
=\frac{\hbar^2}{m_\sigma(\phi)}\int_0^{k^F_\sigma(\phi)}\!dq\;2q\,J_0(qR)\,Y_0(qR).
\end{equation}
The Bessel identity $\int x\,Z_0(x)W_0(x)\,dx=\tfrac{x^2}{2}\big[Z_0W_0+Z_1W_1\big]$, with
$x=qR$, gives
\begin{equation}
\int_0^{k^F_\sigma(\phi)}\!dq\;2q\,J_0(qR)\,Y_0(qR)=k_\sigma^2(\phi)\,\Big[J_0Y_0+J_1Y_1\Big]\big(k^F_\sigma(\phi)R\big),
\end{equation}
where the contribution at the band bottom ($q\to0$) vanishes because
$q^2J_0Y_0\sim q^2\ln q\to0$ and $q^2J_1Y_1\to0$. Using
$k_\sigma^2(\phi)/m_\sigma(\phi)=2(\eF-\sigma\Jp)/\hbar^2$,
\begin{equation}
J_{zz}^{(0)}(\bR)=\frac{\Jc^2M^2}{2\pi\hbar^4}\sum_\sigma \,(\eF-\sigma\Jp)\,
\Big[J_0Y_0+J_1Y_1\Big]\big(k^F_\sigma(\phi)R\big).
\label{eq:Jzz0exact}
\end{equation}

\paragraph{(b) $J_\perp^{(0)}$.} Start from Eq.~\eqref{eq:J0comp} and insert the Green's functions
of Eq.~\eqref{gsigma-epsilon}:
\begin{equation}
\begin{array}{rcl}
G_\up(\varepsilon)\,G_\dn(\varepsilon)
&=&-\dfrac{M^2}{4\hbar^4}\,H_0^{(1)}\big(k_\up(\phi,\varepsilon)R\big)\,H_0^{(1)}\big(k_\dn(\phi,\varepsilon)R\big),
\end{array}
\end{equation}
and hence,
\begin{equation}
J_\perp^{(0)}(\bR)=\frac{\Jc^2M^2}{2\pi\hbar^4}\,{\rm Im}\int_{-|\Jp|}^{\eF}\!d\varepsilon\;
H_0^{(1)}\big(k_\up(\phi,\varepsilon)R\big)\,H_0^{(1)}\big(k_\dn(\phi,\varepsilon)R\big).
\label{eq:Jperp0general}
\end{equation}

\paragraph{(c) No equilibrium DM.} $D_z^{(0)}=0$ for any $\beta$ and $\Jp$, consistent with the
reciprocity and inversion symmetry of the equilibrium state.

\section{First-order in {\bf E}: Non-equillibrium RKKY  }
\paragraph{Lesser correction in k-space.} Start from the momentum-space result and using $A_\sigma(\bk,\varepsilon)=2\pi\,\delta\big(\varepsilon-\varepsilon_\sigma(\bk)\big),$
\begin{align}
\delta G^<_\sigma(\bk,\varepsilon)
&=i\,\big[-f_0'(\varepsilon)\big]\,\big[\dk\cdot\nabla_{\bk}\varepsilon_\sigma(\bk)\big]\,A_\sigma(\bk,\varepsilon)\nonumber\\
&=2\pi i\,\big[-f_0'(\varepsilon)\big]\,\big[\dk\cdot\nabla_{\bk}\varepsilon_\sigma(\bk)\big]\,\delta\big(\varepsilon-\varepsilon_\sigma(\bk)\big),\\
&=-2\pi i\,\big[-f_0'(\varepsilon)\big]\,\dk\cdot\nabla_{\bk}\,\theta\big(\varepsilon-\varepsilon_\sigma(\bk)\big).
\label{eq:dGlk}
\end{align}
\subsection{Fourier transform and integration by parts}
\label{sec:SFT}

The current-induced lesser function in momentum space, Eq.~\mref{eq:gklesser}, contains the
derivative of the step function, $\dk\cdot\nabla_{\bk}\theta\big(\varepsilon-\varepsilon_\sigma(\bk)\big)$,
with $\dk=-e\tau\mathbf E/\hbar$. Its Fourier transform  is evaluated by
integration by parts. Since $\dk$ is constant, the product rule gives
\begin{equation}
e^{i\bk\cdot\bR}\,\dk\cdot\nabla_{\bk}\theta
=\dk\cdot\nabla_{\bk}\big(e^{i\bk\cdot\bR}\,\theta\big)-\theta\,\dk\cdot\nabla_{\bk}e^{i\bk\cdot\bR}.
\label{eq:productrule}
\end{equation}
The first term is a total derivative. By Gauss's theorem its integral reduces to a boundary
term at $|\bk|\to\infty$,
\begin{equation}
\int d^2k\;\dk\cdot\nabla_{\bk}\big(e^{i\bk\cdot\bR}\,\theta\big)
=\oint_{|\bk|\to\infty}dS_{\bk}\,(\dk\cdot\hat{\mathbf n})\,e^{i\bk\cdot\bR}\,
\theta\big(\varepsilon-\varepsilon_\sigma(\bk)\big)=0,
\label{eq:boundary}
\end{equation}
which vanishes because $\theta=0$ outside the Fermi contour. Using
$\nabla_{\bk}e^{i\bk\cdot\bR}=i\bR\,e^{i\bk\cdot\bR}$ in the second term,
\begin{equation}
\int\!\frac{d^2k}{(2\pi)^2}\,e^{i\bk\cdot\bR}\,\dk\cdot\nabla_{\bk}\theta
=-\int\!\frac{d^2k}{(2\pi)^2}\,\theta\;\dk\cdot\nabla_{\bk}e^{i\bk\cdot\bR}
=-i\,(\dk\cdot\bR)\,N_\sigma(\bR,\varepsilon),
\label{eq:ibp}
\end{equation}
where
\begin{equation}
N_\sigma(\bR,\varepsilon)=\int\!\frac{d^2k}{(2\pi)^2}\,e^{i\bk\cdot\bR}\,
\theta\big(\varepsilon-\varepsilon_\sigma(\bk)\big)
\label{eq:Ndef}
\end{equation}
is the real-space transform of the states below $\varepsilon$.
\\
\\
{\bf Calculation of $N_\sigma(\bR,\varepsilon)$ for the elliptic bands.}
\\
We evaluate
\begin{equation}
N_\sigma(\bR,\varepsilon)=\int\!\frac{d^2k}{(2\pi)^2}\,e^{i\bk\cdot\bR}\,\theta\big(\varepsilon-\varepsilon_\sigma(\bk)\big),
\qquad
\varepsilon_\sigma(\bk)=\frac{\hbar^2k_x^2}{2m_{x\sigma}}+\frac{\hbar^2k_y^2}{2m_{y\sigma}}+\sigma\Jp .
\end{equation}
\textit{Assumptions:} (i) $m_{x\sigma},m_{y\sigma}>0$, i.e.\ $|b|<1$, so each band is an
ellipse; (ii) $\varepsilon>\sigma\Jp$, so $\varepsilon$ lies above the band bottom; (iii)
$R\neq0$.

\textit{Step 1: rescale the momenta.} For each spin, substitute
$k_j=\sqrt{m_{j\sigma}}\,p_{\sigma,j}$ ($j=x,y$), with
$\mathbf p_\sigma=(p_{\sigma,x},p_{\sigma,y})$ and $p_\sigma=|\mathbf p_\sigma|$. 
 For an ellipse define three directional quantities will be used later:
\begin{equation}
\begin{gathered}
m_\sigma(\phi)=m_{x\sigma}\cos^2\!\phi+m_{y\sigma}\sin^2\!\phi=\frac{m\,(1-\sigma b\cos2\phi)}{1-b^2},\\
k^F_\sigma(\phi)=\frac{\sqrt{2m_\sigma(\phi)(\eF-\sigma\Jp)}}{\hbar},
\qquad
v_\sigma(\phi)=\frac{\hbar k^F_\sigma(\phi)}{m_\sigma(\phi)} .
\end{gathered}
\label{eq:directional}
\end{equation}
The density-of-states mass is spin independent,
$M=\sqrt{m_{x\sigma}m_{y\sigma}}=m/\sqrt{1-b^2}$. To first order in $\Jp/\eF$
and $b$:
\begin{equation}
k_\sigma\simeq k_F\Big[1-\tfrac{\sigma}{2}\Big(\tfrac{\Jp}{\eF}+b\cos2\phi\Big)\Big],\qquad
v_\sigma\simeq v_F\Big[1-\tfrac{\sigma}{2}\Big(\tfrac{\Jp}{\eF}-b\cos2\phi\Big)\Big] .
\label{eq:firstorder}
\end{equation}
Also,
\begin{equation}
\begin{array}{rcl}
dk_x\,dk_y&=&\sqrt{m_{x\sigma}m_{y\sigma}}\;dp_{\sigma,x}\,dp_{\sigma,y}\equiv M\,d^2p_\sigma,\\
\dfrac{\hbar^2k_x^2}{2m_{x\sigma}}+\dfrac{\hbar^2k_y^2}{2m_{y\sigma}}&=&\dfrac{\hbar^2}{2}\big(p_{\sigma,x}^2+p_{\sigma,y}^2\big)=\dfrac{\hbar^2p_\sigma^2}{2},\\
\bk\cdot\bR&=&\sqrt{m_{x\sigma}}\,p_{\sigma,x}R_x+\sqrt{m_{y\sigma}}\,p_{\sigma,y}R_y=\mathbf p_\sigma\cdot\bm\rho_\sigma,\\
\bm\rho_\sigma&\equiv&\big(\sqrt{m_{x\sigma}}\,R_x,\;\sqrt{m_{y\sigma}}\,R_y\big).
\end{array}
\end{equation}
The step function becomes isotropic in $\mathbf p_\sigma$: the Fermi ellipse is mapped onto a
circle of radius $\kappa_\sigma$,
\begin{equation}
\begin{array}{rcl}
\theta\big(\varepsilon-\varepsilon_\sigma(\bk)\big)&=&\theta\Big(\varepsilon-\sigma\Jp-\dfrac{\hbar^2p_\sigma^2}{2}\Big)
=\theta\big(\kappa_\sigma^2-p_\sigma^2\big)=\theta(\kappa_\sigma-p_\sigma),\\
\kappa_\sigma&=&\dfrac{\sqrt{2(\varepsilon-\sigma\Jp)}}{\hbar},
\end{array}
\end{equation}
where $p_\sigma\ge0$ and $\kappa_\sigma$ is real by assumption (ii). 
Hence
\begin{equation}
N_\sigma(\bR,\varepsilon)=M\int\!\frac{d^2p_\sigma}{(2\pi)^2}\,e^{i\mathbf p_\sigma\cdot\bm\rho_\sigma}\,\theta(\kappa_\sigma-p_\sigma).
\end{equation}
The integrand depends on $\mathbf p_\sigma$ only through
$p_\sigma$ and $\mathbf p_\sigma\cdot\bm\rho_\sigma$, so measure the polar angle $\vartheta_\sigma$ of
$\mathbf p_\sigma$ from the direction of $\bm\rho_\sigma$. With $\rho_\sigma\equiv|\bm\rho_\sigma|$,
\begin{equation}
\begin{array}{rcl}
d^2p_\sigma&=&p_\sigma\,dp_\sigma\,d\vartheta_\sigma,\qquad
\mathbf p_\sigma\cdot\bm\rho_\sigma=p_\sigma\,\rho_\sigma\cos\vartheta_\sigma,\\
N_\sigma(\bR,\varepsilon)&=&\dfrac{M}{(2\pi)^2}\displaystyle\int_0^{\kappa_\sigma}\!p_\sigma\,dp_\sigma\int_0^{2\pi}\!d\vartheta_\sigma\;e^{ip_\sigma\rho_\sigma\cos\vartheta_\sigma}.
\end{array}
\end{equation}
 The integral representation
$J_0(x)=\frac{1}{2\pi}\int_0^{2\pi}d\vartheta\,e^{ix\cos\vartheta}$ gives
\begin{equation}
\begin{array}{rcl}
\displaystyle\int_0^{2\pi}\!d\vartheta_\sigma\;e^{ip_\sigma\rho_\sigma\cos\vartheta_\sigma}&=&2\pi\,J_0(p_\sigma\rho_\sigma),\\
N_\sigma(\bR,\varepsilon)&=&\dfrac{M}{2\pi}\displaystyle\int_0^{\kappa_\sigma}\!p_\sigma\,dp_\sigma\;J_0(p_\sigma\rho_\sigma).
\end{array}
\end{equation}
Substitute $t=p_\sigma\rho_\sigma$ (allowed since $\rho_\sigma>0$
for $R\neq0$), so $p_\sigma\,dp_\sigma=t\,dt/\rho_\sigma^2$:
\begin{equation}
N_\sigma(\bR,\varepsilon)=\frac{M}{2\pi\rho_\sigma^2}\int_0^{\kappa_\sigma\rho_\sigma}\!t\,J_0(t)\,dt .
\end{equation}
The Bessel identity $\frac{d}{dt}\big[t\,J_1(t)\big]=t\,J_0(t)$, together with $0\cdot J_1(0)=0$,
gives $\int_0^x t\,J_0(t)\,dt=x\,J_1(x)$. Therefore
\begin{align}
N_\sigma(\bR,\varepsilon)&=\frac{M}{2\pi\rho_\sigma^2}\,\kappa_\sigma\rho_\sigma\,J_1(\kappa_\sigma\rho_\sigma)
=\frac{M\,\kappa_\sigma}{2\pi\rho_\sigma}\,J_1(\kappa_\sigma\rho_\sigma),\\
a_\sigma(\bR,\varepsilon)&=-2\pi(\dk\cdot\bR)N_\sigma(\bR,\varepsilon)
=2\pi\,(\dk\cdot\bR)\,\dfrac{M\,\kappa_\sigma}{2\pi\rho_\sigma}\,J_1(\kappa_\sigma\rho_\sigma).
\end{align}
With $\bR=R(\cos\phi,\sin\phi)$,
\begin{equation}
\begin{array}{rcl}
\rho_\sigma&=&\sqrt{m_{x\sigma}R_x^2+m_{y\sigma}R_y^2}=R\sqrt{m_\sigma(\phi)},\\
m_\sigma(\phi)&\equiv&m_{x\sigma}\cos^2\phi+m_{y\sigma}\sin^2\phi,\\
\kappa_\sigma\rho_\sigma&=&\dfrac{\sqrt{2m_\sigma(\phi)(\varepsilon-\sigma\Jp)}}{\hbar}\,R\equiv k_\sigma(\phi,\varepsilon)\,R,\\
k_\sigma(\phi,\epsilon)&=&\frac{\sqrt{2m_\sigma(\phi)(\epsilon-\sigma\Jp)}}{\hbar},\\
\dfrac{\kappa_\sigma}{\rho_\sigma}&=&\dfrac{k_\sigma(\phi,\varepsilon)/\sqrt{m_\sigma(\phi)}}{R\sqrt{m_\sigma(\phi)}}=\dfrac{k_\sigma(\phi,\varepsilon)}{m_\sigma(\phi)\,R},\\
\dk\cdot\bR&=&R\,(\dk\cdot\Rh),\\
v_\sigma(\phi,\varepsilon)&=&\hbar k_\sigma(\phi,\varepsilon)/m_\sigma(\phi),
\end{array}
\end{equation}
and,
\begin{equation}
\begin{array}{rcl}
a_\sigma(\bR,\varepsilon)&=&-2\pi\,(\dk\cdot\bR)\,\dfrac{M\,\kappa_\sigma}{2\pi\rho_\sigma}\,J_1(\kappa_\sigma\rho_\sigma)\\
&=&-M\,\dfrac{\kappa_\sigma}{\rho_\sigma}\,(\dk\cdot\bR)\,J_1\big(k_\sigma(\phi,\varepsilon)R\big)\\
&=&-M\,\dfrac{k_\sigma(\phi,\varepsilon)}{m_\sigma(\phi)\,R}\,R\,(\dk\cdot\Rh)\,J_1\big(k_\sigma(\phi,\varepsilon)R\big)\\
&=&-\dfrac{M}{\hbar}\,(\dk\cdot\Rh)\,v_\sigma(\phi,\varepsilon)\,J_1\big(k_\sigma(\phi,\varepsilon)R\big).
\end{array}
\label{eq:a}
\end{equation}
\\
{\bf Expression of $\delta J_{ab}(\bR)$:}
\\
We write,
\begin{equation}
\begin{array}{rcl}
\delta G^<(\bR,\varepsilon)&=&\big[-f_0'(\varepsilon)\big]\,\hat a(\bR,\varepsilon),\\
\hat a(\bR,\varepsilon)&=&{\rm diag}\big(a_\up(\bR,\varepsilon),\,a_\dn(\bR,\varepsilon)\big).
\end{array}
\label{eq:dGlR}
\end{equation}
By Eq.~\eqref{eq:aprops}, each $a_\sigma$ is real and odd in $\bR$. Hence
\begin{equation}
\begin{array}{rcl}
\hat a^\dagger(\bR,\varepsilon)&=&\hat a(\bR,\varepsilon),\\
\hat a(-\bR,\varepsilon)&=&-\hat a(\bR,\varepsilon),\\
\delta G^<(-\bR,\varepsilon)&=&-\big[-f_0'(\varepsilon)\big]\,\hat a(\bR,\varepsilon).
\end{array}
\label{eq:ahatprops}
\end{equation}
\\
 With Eq.~\eqref{eq:ahatprops},
\begin{equation}
\begin{array}{rcl}
{\rm Tr}\big[\sigma_aG^R(\bR,\varepsilon)\,\sigma_b\,\delta G^<(-\bR,\varepsilon)\big]
&=&-\big[-f_0'(\varepsilon)\big]\,Z(\bR,\varepsilon),\\
Z(\bR,\varepsilon)&\equiv&{\rm Tr}\big[\sigma_a\,G^R(\bR,\varepsilon)\,\sigma_b\,\hat a(\bR,\varepsilon)\big].
\end{array}
\end{equation}
Convert the advanced function into a retarded one with:
\begin{equation}
\begin{array}{rcl}
G^A(-\bR,\varepsilon)&=&\big[G^R(\bR,\varepsilon)\big]^\dagger,\\
{\rm Tr}\big[\sigma_a\,\delta G^<(\bR,\varepsilon)\,\sigma_b\,G^A(-\bR,\varepsilon)\big]
&=&\big[-f_0'(\varepsilon)\big]\,{\rm Tr}\big[\sigma_a\,\hat a(\bR,\varepsilon)\,\sigma_b\,G^{R\dagger}(\bR,\varepsilon)\big].
\end{array}
\end{equation}
Using
\begin{equation}
\hat a=\hat a^\dagger,\qquad \sigma_{a,b}=\sigma_{a,b}^\dagger,\qquad
(ABCD)^\dagger=D^\dagger C^\dagger B^\dagger A^\dagger,\qquad
{\rm Tr}\,M^\dagger=({\rm Tr}\,M)^*,
\end{equation}
and cyclicity of the trace,
\begin{equation}
\begin{aligned}
{\rm Tr}\big[\sigma_a\,\hat a\,\sigma_b\,G^{R\dagger}\big]
&={\rm Tr}\Big[\big(G^R\,\sigma_b\,\hat a\,\sigma_a\big)^\dagger\Big]
=\Big\{{\rm Tr}\big[G^R\,\sigma_b\,\hat a\,\sigma_a\big]\Big\}^*\\
&=\Big\{{\rm Tr}\big[\sigma_a\,G^R\,\sigma_b\,\hat a\big]\Big\}^*=Z^* ,
\end{aligned}
\end{equation}
with all functions at $(\bR,\varepsilon)$.
The bracket is,
\begin{equation}
\big[-f_0'(\varepsilon)\big]\big(Z^*-Z\big)=-2i\,\big[-f_0'(\varepsilon)\big]\,{\rm Im}\,Z ,
\qquad
-\frac{i\Jc^2}{2\pi}\cdot(-2i)=-\frac{\Jc^2}{\pi},
\end{equation}
so that
\begin{equation}
\delta J_{ab}(\bR)=-\frac{\Jc^2}{\pi}\int d\varepsilon\,\big[-f_0'(\varepsilon)\big]\,
{\rm Im}\,{\rm Tr}\big[\sigma_a\,G^R(\bR,\varepsilon)\,\sigma_b\,\hat a(\bR,\varepsilon)\big].
\label{eq:dJIm}
\end{equation}
At $T=0$, $-f_0'(\varepsilon)=\delta(\varepsilon-\eF)$, and the energy integral collapses onto the
Fermi surface:
\begin{equation}
\delta J_{ab}(\bR)=-\frac{\Jc^2}{\pi}\,
{\rm Im}\,{\rm Tr}\big[\sigma_a\,G^R(\bR,\eF)\,\sigma_b\,\hat a(\bR)\big],\qquad
\hat a(\bR)\equiv\hat a(\bR,\eF).
\label{eq:dJFS}
\end{equation}
\\
{\bf Calculation at Fermi-energy} 
\\
In \eqref{eq:G} using $H_0^{(1)}=J_0+iY_0$ and
$k^F_\sigma(\phi)\equiv k_\sigma(\phi,\eF)$,
\begin{equation}
G^R_\sigma(\bR,\eF)=-\frac{iM}{2\hbar^2}\Big[J_0(k^F_\sigma(\phi) R)+i\,Y_0(k^F_\sigma(\phi) R)\Big]
=\frac{M}{2\hbar^2}\,Y_0(k^F_\sigma(\phi) R)-i\,\frac{M}{2\hbar^2}\,J_0(k^F_\sigma(\phi) R),
\end{equation}
that is,
\begin{equation}
\begin{array}{rcl}
g_\sigma&\equiv&G^R_\sigma(\bR,\eF)=g'_\sigma+i\,g''_\sigma,\\
g'_\sigma&=&\dfrac{M}{2\hbar^2}\,Y_0(k^F_\sigma(\phi) R),\\
g''_\sigma&=&-\dfrac{M}{2\hbar^2}\,J_0(k^F_\sigma(\phi) R).
\end{array}
\label{eq:gparts}
\end{equation}

\textit{Check.} The imaginary part follows independently from the spectral function, which
needs no contour. With $\delta(E-\hbar^2p^2/2)=\delta(p-\kappa_\sigma)/(\hbar^2\kappa_\sigma)$,
\begin{equation}
A_\sigma(\bR,\eF)=M\int_0^\infty\!dp\,p\,J_0(p\rho_\sigma)\,2\pi\,\frac{\delta(p-\kappa_\sigma)}{2\pi\,\hbar^2\kappa_\sigma}\cdot2\pi\Big/2\pi
=\frac{M}{\hbar^2}\,J_0(k^F_\sigma(\phi) R)=-2\,g''_\sigma ,
\end{equation}
consistent with $A_\sigma=-2\,{\rm Im}\,G^R_\sigma$ for even bands. Only the real part
$g'_\sigma\propto Y_0$, the reactive part that produces $D_z^{(1)}$, needs the contour.
Define the Fermi-energy quantities
\begin{equation}
\begin{array}{rcl}
g_\sigma&\equiv&G^R_\sigma(\bR,\eF)=g'_\sigma+i\,g''_\sigma,\\
a_\sigma&\equiv&a_\sigma(\bR,\eF)\in\mathbb R .
\end{array}
\end{equation}
\\
\textit{calculate trace.} Write the trace of the product of four $2\times2$ matrices with explicit spin
indices:
\begin{equation}
{\rm Tr}\big[\sigma_a\,G^R\,\sigma_b\,\hat a\big]
=\sum_{\sigma\sigma'\sigma''\sigma'''}\big(\sigma_a\big)_{\sigma\sigma'}\,\big(G^R\big)_{\sigma'\sigma''}\,
\big(\sigma_b\big)_{\sigma''\sigma'''}\,\big(\hat a\big)_{\sigma'''\sigma}.
\end{equation}
Both $G^R$ and $\hat a$ are diagonal in spin,
\begin{equation}
\begin{array}{rcl}
\big(G^R\big)_{\sigma'\sigma''}&=&\delta_{\sigma'\sigma''}\,g_{\sigma'},\\
\big(\hat a\big)_{\sigma'''\sigma}&=&\delta_{\sigma'''\sigma}\,a_\sigma .
\end{array}
\end{equation}
Inserting these, 
\begin{equation}
\begin{array}{rcl}
{\rm Tr}\big[\sigma_a\,G^R\,\sigma_b\,\hat a\big]
&=&\displaystyle\sum_{\sigma\sigma'\sigma''\sigma'''}\sigma^a_{\sigma\sigma'}\,\delta_{\sigma'\sigma''}\,g_{\sigma'}\,
\sigma^b_{\sigma''\sigma'''}\,\delta_{\sigma'''\sigma}\,a_\sigma\\
&=&\displaystyle\sum_{\sigma\sigma'}\sigma^a_{\sigma\sigma'}\,g_{\sigma'}\,\sigma^b_{\sigma'\sigma}\,a_\sigma\\
&=&\displaystyle\sum_{\sigma\sigma'}\underbrace{\sigma^a_{\sigma\sigma'}\,\sigma^b_{\sigma'\sigma}}_{c^{ab}_{\sigma\sigma'}}\,g_{\sigma'}\,a_\sigma ,
\end{array}
\end{equation}
where in the last line the scalars $g_{\sigma'}$ and $a_\sigma$ were moved to the right. Hence
\begin{equation}
\begin{array}{rcl}
{\rm Tr}\big[\sigma_a\,G^R\,\sigma_b\,\hat a\big]&=&\displaystyle\sum_{\sigma\sigma'}c^{ab}_{\sigma\sigma'}\,g_{\sigma'}\,a_\sigma ,\\
c^{ab}_{\sigma\sigma'}&=&\sigma^a_{\sigma\sigma'}\,\sigma^b_{\sigma'\sigma}.
\end{array}
\end{equation}
The outer index $\sigma$ labels the diagonal element and goes with $\hat a$. The inner index $\sigma'$ is summed in the
matrix product and goes with $G^R$.
The coefficients $c^{ab}_{\sigma\sigma'}$ are listed in Table~\ref{tab:cab}. Out of 24 only 8 survive.
\begin{table}
\centering
\caption{Spin-channel coefficients $c^{ab}_{\sigma\sigma'}=\sigma^a_{\sigma\sigma'}\sigma^b_{\sigma'\sigma}$.}
\label{tab:cab}
\begin{tabular}{lcccc}
\toprule
$(a,b)$ & $c_{\up\up}$ & $c_{\dn\dn}$ & $c_{\up\dn}$ & $c_{\dn\up}$\\
\midrule
$(x,x)$ & 0 & 0 & $1$ & $1$\\
$(y,y)$ & 0 & 0 & $(-i)(i)=1$ & $(i)(-i)=1$\\
$(z,z)$ & 1 & 1 & 0 & 0\\
$(x,y)$ & 0 & 0 & $(1)(i)=i$ & $(1)(-i)=-i$\\
$(y,x)$ & 0 & 0 & $-i$ & $i$\\
$(x,z)$, $(z,x)$, $(y,z)$, $(z,y)$ & 0 & 0 & 0 & 0\\
\bottomrule
\end{tabular}
\end{table}
\noindent Reading off the table, the three independent traces are
\begin{equation}
\begin{array}{rcl}
{\rm Tr}\big[\sigma_x\,G^R\,\sigma_x\,\hat a\big]&=&g_\dn\,a_\up+g_\up\,a_\dn,\\
{\rm Tr}\big[\sigma_z\,G^R\,\sigma_z\,\hat a\big]&=&g_\up\,a_\up+g_\dn\,a_\dn,\\
{\rm Tr}\big[\sigma_x\,G^R\,\sigma_y\,\hat a\big]&=&i\,\big(g_\dn\,a_\up-g_\up\,a_\dn\big).
\end{array}
\end{equation}
Since $a_\sigma$ is real,
\begin{equation}
\begin{array}{rcl}
{\rm Im}\big(g_\sigma\,a_{\sigma'}\big)&=&g''_\sigma\,a_{\sigma'},\\
{\rm Im}\big(i\,g_\sigma\,a_{\sigma'}\big)&=&g'_\sigma\,a_{\sigma'}.
\end{array}
\end{equation}
Equation~\eqref{eq:dJFS} then gives
\begin{equation}
\begin{array}{rcl}
\delta J_\perp\equiv\delta J_{xx}=\delta J_{yy}&=&-\dfrac{\Jc^2}{\pi}\big(a_\up\,g''_\dn+a_\dn\,g''_\up\big),\\
\delta J_{zz}&=&-\dfrac{\Jc^2}{\pi}\big(a_\up\,g''_\up+a_\dn\,g''_\dn\big),\\
D_z^{(1)}=-\delta J_{xy}=\delta J_{yx}&=&\dfrac{\Jc^2}{\pi}\big(a_\up\,g'_\dn-a_\dn\,g'_\up\big),
\end{array}
\label{eq:dJcomp}
\end{equation}
where $D_z=-J_{xy}$ from Eq.~\eqref{eq:J0comp} was used. 

Order-of-magnitude estimates for two representative platforms are given in Table~\ref{tab:magnitude}. Since
$\delta k/k_F=v_d/v_F$, the current-induced couplings scale with the drift velocity of the
carriers relative to the Fermi velocity. In metallic altermagnets the high carrier density keeps
$v_d/v_F\sim10^{-4}$, so the effects are weak; in low-density, high-mobility 2D layers $v_d/v_F$
is two orders of magnitude larger and the chiral term reaches a few percent of $J^{(0)}$, within
the ballistic ($R<\ell$) and linear ($\delta k\,R\ll1$) regime.

\begin{table}[t]
\caption{Order-of-magnitude estimates for a bond along the
current ($\phi=0$), with $k_FR=20$ and $|\Jp/\eF-b|=0.2$. The drift velocity is
$v_d=j/ne$ (film) or $v_d=K/ne$ (sheet current density $K$), and $\delta k/k_F=v_d/v_F$. The $|\delta J|/|J^{(0)}|$ stands for $|\delta J_{zz}|/|J_{zz}^{(0)}|\simeq|\delta J_\perp|/|J_\perp^{(0)}|$.}
\label{tab:magnitude}
\centering
\small
\begin{tabular}{llcccc}
\toprule
Platform & Inputs & $v_d/v_F$ & $\delta k\,R$ & $|\delta J|/|J^{(0)}|$ & $|\delta D_z|/|J^{(0)}|$\\
\midrule
\parbox[t]{2.6cm}{\raggedright Metallic altermagnetic film} &
\parbox[t]{6.4cm}{\raggedright $n\approx10^{28}\,{\rm m^{-3}}$, $m^*\approx0.8\,m_e$ ($v_F\approx1.0\times10^{6}$\,m/s), $j=10^{11}\,{\rm A/m^2}$; $R\approx3$\,nm}
 & $6.2\times10^{-5}$ & $1.2\times10^{-3}$ & $2.5\times10^{-3}$ & $2.5\times10^{-4}$\\[18pt]
\parbox[t]{2.6cm}{\raggedright Low-density gated 2D layer} &
\parbox[t]{6.4cm}{\raggedright $n\approx10^{16}\,{\rm m^{-2}}$, $m^*\approx0.1\,m_e$ ($v_F\approx2.9\times10^{5}$\,m/s), $K=5$\,A/m; $R\approx80$\,nm}
 & $1.1\times10^{-2}$ & $0.22$ & $0.43$ & $0.04$\\[18pt]
\bottomrule
\end{tabular}
\end{table}
\section{Numerical checks}
\label{sec:Snumchecks}

\begin{figure}
\centering
\includegraphics[width=0.55\textwidth]{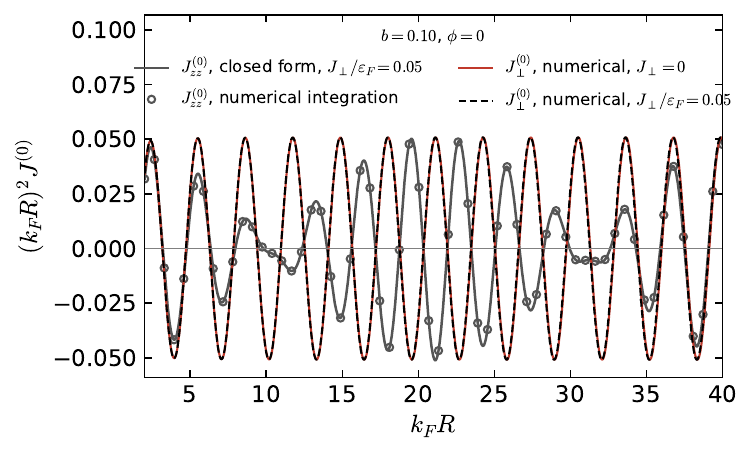}\hfill
\includegraphics[width=0.43\textwidth]{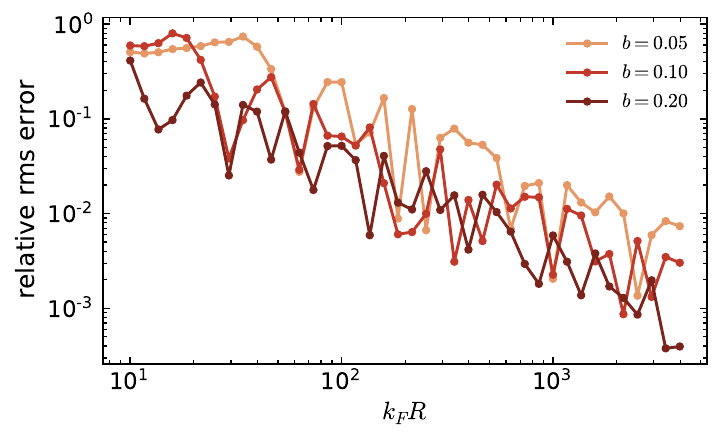}
\caption{(a)~Equilibrium couplings along $\phi=0$ ($b=0.10$): $J_{zz}^{(0)}$ in closed form
(Eq.~\mref{eq:Jzz0exact}) for $\Jp/\eF=0.05$ (line), compared with direct numerical integration
of Eq.~\mref{eq:J0GF} (symbols), and $J_\perp^{(0)}$ from numerical integration of
Eq.~\mref{eq:Jperp0general} for $\Jp=0$ and $\Jp/\eF=0.05$. (b)~Relative rms error of the
long-distance form of $\delta D_z$ (Eq.~\mref{eq:Dasym}) with respect to the exact result
(Eq.~\mref{eq:Dexact}), as a function of $k_FR$ for $b=0.05$, $0.10$, and $0.20$ ($\Jp=0$,
$\phi=0$).}
\label{fig:checks}
\end{figure}

Figure~\ref{fig:checks} summarizes two checks of the numerics. The closed-form $J_{zz}^{(0)}$
agrees with direct numerical energy integration of Eq.~\mref{eq:J0GF} to about $10^{-9}$
(Fig.~\ref{fig:checks}(a)), which confirms the real-space Green's function
(Eq.~\mref{eq:G}). The equilibrium $J_\perp^{(0)}$ has no closed form at finite $\Jp$ and is
therefore computed by numerical integration in all figures. The relative error of the
long-distance form of $\delta D_z$ falls below $10\%$ only beyond $k_FR\approx130$, $65$, and
$35$ for $b=0.05$, $0.10$, and $0.20$, respectively, and below $1\%$ beyond $k_FR\approx750$ for
$b=0.10$ (Fig.~\ref{fig:checks}(b)). The quantitative statements in the main text therefore rely
on the exact Bessel-function expressions.

\end{document}